\documentstyle[prl,aps,multicol,psfig]{revtex}
\tighten
\begin{document}
\twocolumn
\draft
\preprint{\today}
\title{Spectral Statistics in Chiral-Orthogonal Disordered Systems}
\author{S. N. Evangelou$^{1,2*}$ and D.E. Katsanos$^1$ }
\address
{$^1$ Department of Physics, University of Ioannina, Ioannina
45110, Greece 
\\
$^2$ Department of Physics, University of Lancaster,
Lancaster SW1 2BZ, UK} 
\maketitle
\date{\today}

\begin{abstract}

We describe the singularities in the averaged density of states 
and the corresponding statistics of the energy levels in two- (2D)
and three-dimensional (3D) chiral symmetric and time-reversal invariant
disordered systems, realized in bipartite 
lattices with real off-diagonal disorder.
For off-diagonal disorder of zero mean we obtain a singular density of states
in 2D which becomes much less pronounced in 3D, while the level-statistics 
can be described by semi-Poisson distribution with mostly critical
fractal states in 2D and Wigner surmise with mostly delocalized 
states in 3D. 
For logarithmic off-diagonal disorder of large strength 
we find indistinguishable 
behavior from ordinary disorder with strong localization in any dimension
but in addition one-dimensional $1/|E|$ Dyson-like asymptotic spectral 
singularities. 
The off-diagonal disorder is also shown to enhance the propagation of
two interacting particles similarly to systems with diagonal disorder.
Although disordered models with chiral symmetry differ from 
non-chiral ones due to the presence of spectral singularities, 
both share the same qualitative localization properties except 
at the chiral symmetry point $E=0$ which is critical.

\end{abstract}


\narrowtext
\section{INTRODUCTION}

\medskip
In the last couple of decades the effect of various symmetries 
in the scaling theory \cite{1} of Anderson localization 
has been explored, both theoretically and experimentally,
leading to fundamental results \cite{2}. 
For example, in two-dimensional (2D) disordered systems breaking 
of the symmetry under time-reversal leads to the Quantum Hall Effect 
where the presence of magnetic field is responsible for critical 
delocalized states at the center of each Landau band which carry 
the Hall current \cite{3}, while breaking of symmetry under spin-rotation 
due to spin-orbit coupling is believed to lead 
to a metallic phase even in 2D \cite{4}.  
Recently, strong interest has focused on other basic symmetries, 
such as the chiral or particle-hole symmetry.
In addition to the three standard universality classes 
(orthogonal, unitary and symplectic) the presence of chiral symmetry 
gives three extra chiral ensembles \cite{5,6}, while 
four similar but distinct ensembles are realized for 
particle-hole symmetry in normal metal-superconducting 
systems, increasing their total number to ten \cite{7}. 
The prerequisite for chiral symmetry is the presence 
of bipartite lattice structure with two interconnected sublattices and
the simplest case where it can be realized in disordered systems
is real random hopping between nearest-neighbours \cite{8}. 
This is also known as off-diagonal disorder and it is worth studying 
the spectral and localization properties of such
systems in order to see the effect of chirality in  2D, 3D. 
The presence of chiral symmetry in disordered systems 
somehow simplifies both analytical and numerical calculations so that 
more is known about Anderson localization for  the chiral 
symmerty classes in one, quasi-one and two dimensions 
illustrated for quasi-one dimension in \cite{9}.
In the rest of the paper  we will rely on numerical tools 
in order to explore quantities that have not been computed 
analytically so far and also for quantities where the analytical 
predictions are competing, such as the 2D density of states.

\medskip
The chiral symmetry is present in simple nearest-neighbor models
with off-diagonal disorder  defined on bipartite lattices 
which consist of two sublattices $A$ and $B$, one connected to the other 
by random bonds \cite{8}. This geometric kind of symmetry, 
where the randomness strictly 
connects sites of two different sublattices, is different from 
other intrinsic symmetries of the Hamiltonian such 
as time-reversal and spin-rotation. It can also appear as 
an electron-hole transformation in half filled many body 
systems \cite{7,10}. It is well-known that a consequence 
of the chiral symmetry is a  distinct special point in the 
energy spectrum, for off-diagonal 
disorder in square and cubic lattices this is the band center $E=0$ 
with the eigenvalues making up symmetric energy pairs $(E,-E)$ 
around the zero mode $E=0$. For an early observation of localized
chiral zero-modes see Ref. \cite{11}.
For finite lattices the special chiral energy eigenvalue $E=0$ 
exists for any disorder configuration but only when the total 
number of sites is odd. Moreover, the
corresponding $E=0$ wavefunction is non-zero in one
sublattice only and can be easily constructed \cite{12,13}. 
In fact, for any bipartite lattice with off-diagonal disorder 
only and $n_A$ ($n_{B} \leq  n_{A}$) sites belonging to sublattice $A$ 
($B$) the number of linearly independent $E=0$ states
is exactly $n_{A}-n_{B}$, with their amplitudes 
being non-zero on the $A$ sublattice and exactly zero on the $B$ sublattice.
The chiral $E=0$ state turns out to be neither localized nor extended 
but critical in all dimensions,  also being very sensitive to the 
choice of boundary conditions \cite{14}, reminiscent of critical states 
at the mobility edge of the Anderson transition \cite{15}. 
The main issue discussed in this paper 
is the singular behavior of the density of states in 2D, 3D,
when approaching the special chiral symmetry energy  $E=0$ 
where the localization length also diverges \cite{8,16}.

\medskip
The interest in this problem has been recently revived, 
due to the related problem of massless  Dirac fermions in a random gauge 
fields in \cite{17} followed by  \cite{18,19,20} and also the Bogoliubov-de Gennes 
(BdG) Hamiltonians \cite{21} which describe superconducting quasiparticles 
in mean-field theory. 
The Dirac fermions have very different zero disorder limit from
the off-diagonal disorder case. In the tight-binding 
approximation the 2D fermions have a zero energy Fermi surface with many points
forming a square while the Dirac fermions  have a Fermi surface 
with four points only. 
In the former case the pure 2D density of states at $E=0$ 
has the well-known $\log$-type van Hove singularity while in the Dirac case 
the density of states approaches zero at the band center \cite{21}.

\medskip
First, let us summarize what is known about the spectral and wavefunction
anomalies of one-electron spectra in the presence of off-diagonal disorder 
in bipartite lattices described by real symmetric random Hamiltonians.
The 1D random hopping problem in a bipartite lattice has a chiral spectrum 
with a  strong  $1/|E \ln^{3} |E||$ leading Dyson singularity of the 
density of states 
as $E$ approaches zero, which is the first result obtained in 
disordered systems \cite{22}, 
even before the Anderson theory of localization was proposed \cite{23}.
The corresponding states of the energy spectrum are Anderson localized
with their wavefunction amplitude decaying exponentially as 
$|\psi_{E\ne 0}(r)|\sim e^ {-\gamma r}$, where $r$ is the distance 
and $\gamma$ the inverse localization length which depends on the disorder. 
This is not true at the  chiral symmetry zero mode $E=0$,
where the corresponding wavefunction amplitude which lies in one sublattice
decays slower than exponential, as $|\psi_{E=0}(r)|\sim e^{-\gamma \sqrt{r}}$ 
at distance $r$ from its maximum where $\gamma$ also varies with disorder. 
This result for the typical $E=0$ wavefunction can be easily derived
since its log-amplitude $\ln |\psi_{E=0}(r)|$ at site $r$ 
executes a random-walk 
in space \cite{8} leading to the exponential square root decay.  
In higher-dimensional, bipartite systems such as the square 2D 
or cubic 3D lattices the $E=0$ wavefunction was shown to display 
multifractal fluctuations with many scattered peaks and a disorder-dependent
fractal dimension \cite{13}. 
Moreover, an appropriate correlation function  behaves logarithmically 
which roughly implies an overall power-law decay $|\psi_{E=0}(r)| 
\sim r^{-\eta}$ of the peak heights from the maximum peak
and disorder dependent-$\eta$. It must be understood that 
such a power-law description is only a simplified picture for 
an approximate decaying character of the chiral $E=0$ 
critical state.
For chiral systems the density of states at the band center
does not diverge  only in 1D.  In the absence of time-reversal 
symmetry on a 2D bipartite lattice for weak disorder
in the vicinity of the band center the divergence 
$\rho_{2D}(E)\sim |E|^{-1} \exp \left(-c \left(\ln 
{\frac {1}{|E|}}\right)^\kappa \right)$ 
was predicted by Gade \cite{5} with a constant $c$ 
and a universal exponent $\kappa=1/2$. 
If universality applies this singularity should also hold for
the chiral-orthogonal ensemble.
For strong disorder Mortunich, Damle and Huse \cite {24} argued that
the exponent $\kappa=1/2$ does not apply for a typical density of states 
in the chiral orthogonal universality class and predict $\kappa=2/3$ instead.
In \cite{25} the density of states for the $\pi$-flux phase 
is discussed via a weak disorder approach. The analytical 
work of \cite{26} gives a wealth of exact results
for the $\pi$-flux phase weakly perturbed by
real nearest-neighbour hoppings. 
Recent work  \cite{27} for random Dirac fermions instead of vanishing 
density of states shows an upturn close to $E=0$, in agreement with the
diverging dos of Gade \cite{5}. The multifractal analysis of \cite{28}
is compared to the analytical results of \cite{20}.

\medskip
In this paper we shall investigate the consequences of the chiral symmetry 
for real symmetric random matrices in the orthogonal universality class 
by studying the spectrum of 2D, 3D square and cubic tight binding lattices 
with off-diagonal disorder. In the studied systems apart from chirality 
the symmetries of time-reversal and spin-rotation are preserved as well. 
We  shall determine  the midband singularities in the density of states 
$\rho(E)$ and from the nearest-level-statistics
the localization properties of the corresponding eigenstates. 
Our results demonstrate the effect of chirality for  the unique
off-diagonal disorder of mean zero: 
In 2D the nearest level-spacing distribution function $P(S)$
turns out to be intermediate between Wigner and Poisson roughly 
described by the so-called semi-Poisson curve \cite{15} 
(for small-spacing $S$ shows Wigner-like linear increase $\propto S$ and  
for large-spacing $S$ shows Poisson-like exponential decrease $\propto 
exp(-S)$), which indicates the presence of critical states in finite 
2D disordered systems. 
In 3D our results for the eigenvalue spacing fluctuations 
with zero mean off-diagonal disorder are very close to the Wigner surmise
which corresponds to extended states. For logarithmic off-diagonal 
disorder which permits its strength to vary we see
for very strong values that the distributions both 
in 2D, 3D become Poisson-like independently of the dimension 
and the states localize 
similarly to states of ordinary disordered systems with
broken chiral symmetry. 
Apart from clarifying the effect of chiral symmetry for non-interacting 
electron in disordered media our aim is also to extend our calculations 
to more than one electron by including Hubbard electron-electron  
interactions together with off-diagonal disorder, 
via the simplest possible two-electron Hamiltonian. This is done in order 
to discuss a recent conjecture concerning the decisive role of chiral symmetry 
for quantum transport properties in the presence of both disorder and interactions 
at half-filling, where enhancement of the conductivity with increasing disorder
was recently obtained only for non-chiral systems at half-filling
\cite{29}.

\medskip
We present a numerical study of the spectrum and its 
fluctuations in squared and cubic bipartite lattices with
disorder in the nearest-neighbor hoppings. We summarize our main 
questions: 
(i) What is the nature of the spectral singularities 
in the presence of chiral symmetry?
(ii) What is the corresponding level-statistics and the localization 
behavior for disordered systems when chiral symmetry is preserved?
(iii) From the established spectral and localization behavior for systems 
with chiral or broken chiral symmetry can we conclude whether all states 
remain localized in 2D?
(iv) How is the electron-electron interaction affected
by off-diagonal disorder for two particles only?
Apart from the theoretical interest in answering the above questions 
we  emphasize that our study of localization with off-diagonal disorder 
can have a lot of applications. In particular, the presence or not 
of the chiral symmetry could be useful for understanding various properties 
of several realistic 2D systems, such as current-currying states in the 
quantum Hall effect \cite{3}, quasiparticles in dirty superconductors 
\cite{27}, states in semiconductor quantum wells, high mobility 
silicon MOSFETs \cite{30}, etc.

\section{THE  DISORDERED HAMILTONIAN WITH CHIRAL SYMMETRY}

\medskip
We consider a tight-binding model Hamiltonian defined in 
bipartite lattices with random nearest-neighbor hoppings
of the off-diagonal block symmetric structure 
\begin{equation}
H=\sum_{\langle ij \rangle }(t_{ij}c^{\dag }_ic_j+\text{H.c.})
= \left(\begin{array}{cc} 
    0            & H_{A,B} \\
    H^{+}_{A,B}  & 0
\end{array}                \right),
\end{equation}
where the sum is taken over all bonds 
${\langle ij \rangle }$ where
$i,j$ denote nearest neighbor lattice sites
which belong to different sublattices, $c_i$ is the annihilation
operator of an electron on site $i$ and $t_{ij}$ the nearest-neighbor 
hopping integrals which are real independent random variables satisfying 
a box probability distribution with mean zero and arbitrary variance. 
This is equivalent to choosing random hoppings $t_{i,j}$  
from a box distribution $P(t_{ij}) = \frac{1}{w},
\text{  for  }-\frac{w}{2} \leq t_{ij} \leq  \frac{w}{2}$ 
of  zero mean and width $w$. The scaling $t_{i,j}/w$ 
is sufficient to make the disorder independent of $w$, 
due to absence of other energy scale in the Hamiltonian. 
Thus, by increasing $w$ the  energy band becomes wider 
since it amounts to a rigid rescaling of the eigenvalues only 
(the eigenstates remain the same for any $w$) and
this is a unique form of ${\it {zero}}$ ${\it {mean}}$ off-diagonal disorder 
which allows certain comparison with analytical results to be made.
The alternative choice is disorder of mean zero 
and width $W$ but for the $\ln t_{i,j}$. 
This ${\it logarithmic}$ disorder distribution is $W$-dependent, 
can be broad becoming arbitrarily strong
and guarantees always positive hoppings $t_{i,j}$. 
It is commonly used in 1D where the real hoppings $t_{i,j}$ 
chosen from a logarithmic range must be always positive 
and can be widely varied with typical $W$, which is  
believed to be a good measure of off-diagonal 
disorder more generally.
We emphasize that in our study the random distribution of the hoppings 
$t_{ij}$ is the only disorder ingredient in the Hamiltonian $H$, 
since we completely ignore diagonal disorder by setting all site energies 
equal to zero. The diagonal disorder, somehow, connects each site
to itself breaking the chiral symmetry. In our studies we consider 
finite  $L\times L$ square and $L\times L \times L$ cubic
lattices with various choices of boundary conditions, taking care to
preserve the bipartite structure and chirality of $H$. 

\medskip
In Eq. (1) the matrix $H_{A,B}$ contains matrix elements 
which connect the $A$, $B$ interconnected sublattices 
defining the two $A$, $B$
bases and due to chiral symmetry the Hamiltonian transforms as 
\begin{equation}
H=-\sigma_{3} H \sigma_{3}, \;\;\;
 \sigma_{3}=\left(\begin{array}{cc} 
    1            & 0 \\
    0            & -1
\end{array}                \right),
\end{equation}
which means anticommuting $H$ and $\sigma_{3}$. One notices that 
$\sigma_{3}$ behaves like the $\gamma_{5}$ familiar from applications 
of gauge theories \cite{17}. Moreover, the presence of chiral symmetry 
allows to reduce the size of the corresponding matrices to half by 
diagonalizing $H^{2}$ instead of $H$ obtaining the  eigenvalues squared.
Due to the interconnected by the random hoppings sublattice structure 
of the studied lattices \cite{12} the eigenvalues appear in pairs $E, -E$ 
around $E=0$ so that the spectral density is strictly 
an even function $\rho(E)=\rho(-E)$ of the energy $E$. 
The eigenvalue pairs $E, -E$ around $E=0$ 
have simply related eigenstates $\psi_{E}(\vec{r})=\sum_{n,l,...} 
\psi_{n,l,...} |n,l,... \rangle$ and $\psi_{-E}(\vec{r})
=\sum_{n,l,...} (-1)^{n+l+...} \psi_{n,l,...} |n,l,... \rangle$
since $H\psi=E\psi$ from Eq. (2) gives $H \sigma_{3}\psi=
-E\sigma_{3}\psi$ so that the unitary transformation $\sigma_{3}$ 
changes the wavefunction amplitude of the $-E$ state on every other site,
say on the sites of sublattice $B$, keeping the amplitudes of the $-E$ 
state the same as that of $E$ on sublattice $A$. 
Moreover, the $E=0$ state when it exists (for odd total number of sites)
has amplitude only on $A$ sublattice and is strictly zero on $B$ sublattice.

\medskip
In order to measure the anomalies in the spectral density 
and their corresponding fluctuations we obtained the eigenvalues 
of finite Hamiltonian matrices of size $L^{d}\times L^{d}$, $d=2,3$  
for 2D, 3D, respectively.
The very sparse matrices are obtained from Eq. (1) by
distributing either $t_{i,j}$ uniquely  or $\ln t_{i,j}$, 
with various strengths $W$ to cover a wider range 
of off-diagonal disorder.
We have used  standard diagonalization algorithms, including Gaussian
elimination for computing the integrated density of states (idos)
by employing eigenvalue counting theorems
at fixed $E$ \cite{31} and the Lanczos algorithm for computing eigenvalues 
within an energy range \cite{32}. Moreover, we have considered all 
possible boundary conditions \cite{15}, such as periodic in all directions, 
periodic in one and hard wall in the other or hard wall in all directions, 
always preserving the chiral symmetry. 
It must be also noted that crucial even-odd size effects \cite{9,33} 
may arise in this problem, 
for example, with periodic boundary conditions the sites of the 
$A$ sublattice are connected to  the sites of the $B$ sublattice 
preserving chirality only for even linear size $L$. 
The periodic boundary conditions join sites which  
no longer belong to different sublattices for odd $L$  
and the chiral symmetry is broken.
For hard wall boundary conditions no such even-odd effects arise. 
We should also remember that the zero mode $E=0$ for chiral
systems exists  for odd sizes only. More details of the numerical 
methods can be found in Refs. \cite{31,32}.

\section{ZERO MEAN OFF-DIAGONAL DISORDER}

\subsection{The singularities in the dos}

\medskip
In one dimension (1D) we cannot have zero mean
off-diagonal disorder due to the finite probability of finding a zero bond 
which breaks the chain. 
Outside one dimension for this kind of disorder some rigorous results 
are known, such as the Gade \cite{5} and Motrunich et al \cite{24} 
singularities. Earlier approximate results were derived with a
${\frac {1}{N}}$ - expansion \cite{34} since
the studied model is an example of the ensemble introduced
in Ref. \cite{10}. The extrapolation of the one-loop results 
near $E=0$ \cite{32} gives a logarithmic singularity in 2D dos
\begin{equation}
\rho_{2D}(E)=0.299577...+{\frac{1}{2\pi^2}}\ln {\frac {1}{|E|}}+ O(|E|),
\end{equation}
similar
to that of  the pure system where the constant term 0.299577 is replaced 
by ${\frac {1}{2\pi^2}}(\ln16+(1/2)\ln2) \approx 0.1580186$ \cite{32,34}.
In the two-loop approximation the situation is less clear, 
for example, the corresponding term is divergent, proportional
to $(\ln {\frac {1}{|E|}})^2$, 
which might imply a power-law $\rho_{2D}(E)\propto |E|^{-\phi}$  
spectral singularity.
The small $|E|$ expression obtained in 2D by Gade \cite{5} 
combines the strong $1/|E|$ divergence (Dyson-type) with a slowly decaying 
exponential of the logarithm of $E$ to the power $\kappa$, 
written as their product
\begin{equation}
\rho_{2D}(E)={\frac {c_{1}}{|E|}} \exp \left(-c_{2} \left(\ln 
{\frac {1}{|E|}}\right)^{\kappa}\right), \;\;\; \kappa=1/2,
\end{equation}
where $c_{1}, c_{2}$ are model-dependent constants. The improved 
strong disorder expression obtained by Motrunich ${\it {etal}}$ 
\cite{24} involves a different exponent for the logarithm $\kappa=2/3$.
In 3D the one-loop result gives a square-root singularity which remains
for higher loops implying finite dos at $E=0$ with
\begin{equation}
\rho_{3D}(E)=0.400494...-{\frac {3\sqrt{3}}{2\pi^2}} \sqrt{|E|} + O(|E|).
\end{equation}

\medskip
In Fig. 1 numerical results are presented for the  averaged integrated dos 
(idos) ${\cal N}(E)={\int_{0}^{E}{dE' \rho(E')}}$ and the dos $\rho(E)$
close to $E=0$. 
The random hoppings $t_{i,j}$ are uniformly distributed for example
within $[-1/2,+1/2]$ since, of course, this specific range is irrelevant 
for localization. In order to determine 
the spectral singularities two possible fits are tried:

${\it {(i)}}$ A power-law fit for the idos ${\cal {N}}(E)\propto |E|^{1-\phi}$
which implies for the dos the power-law divergence $\rho(E)\propto |E|^{-\phi}$.
In Fig. 1(a) numerical data for the idos ${\cal {N}}(E)$ which are obtained 
from eigenvalue-counting techniques allow to compute the exponent $\phi$ 
for squared and cubic lattice with lattice sizes up to $600\times 600$ 
in 2D and $30\times 30\times 30$ in 3D. 
In the log-log plots the data are seen to collapse to power-laws which 
define a small exponent $\phi\approx 0.25$ in 2D and $\phi\approx 0.07$ in 3D.
These weak power-law divergencies of the dos with off-diagonal disorder 
are certainly different from the expected log-type singularities 
of non-random systems in pure 2D and the constant dos in pure 3D (see Fig. 1).

${\it {(ii)}}$ We have also tried  an alternative fit of the data in Fig. 1(b)
by plotting directly $\rho(E)$ vs. $E$ in a semi-log plot. 
In 2D the data can be seen to be reasonably explained
by the analytical form of Gade \cite{5} and Motrunich ${\it {etal}}$ 
\cite{24}.
The fit of Gade with $\kappa=1/2$ gave $c_{1}\approx 21.5, c_{2}\approx 3.77$ 
and the fit of Motrunich ${\it {etal}}$ with $\kappa=2/3$ gave
$c_{1}\approx 2.61, c_{2}\approx 2.14$. We have also tried another 
fit of the data which gave a slightly weaker power-law in the exponential
with the value $\kappa \approx 0.91$ and constants $c_{1}\approx 0.47, c_{2}\approx 1.04$. 
We also see that the one-loop result of Eq. (5) in 3D 
is clearly inappropriate to fit the small-$|E|$ data.  Other fits
can be also tried such as the fit of Ref. \cite{35} which instead
of $1/|E|$ from Eq. (4) involves the square-root $1/\sqrt{|E|}$ 
leading divergence.
The above fits (i), (ii) verify that the presence of zero mean 
off-diagonal disorder leads to a weakly singular dos at least in 2D 
and a much weaker singularity, 
or a simple increase of the dos close to $E=0$, in 3D.

\medskip 
%
%
\centerline{\psfig{figure=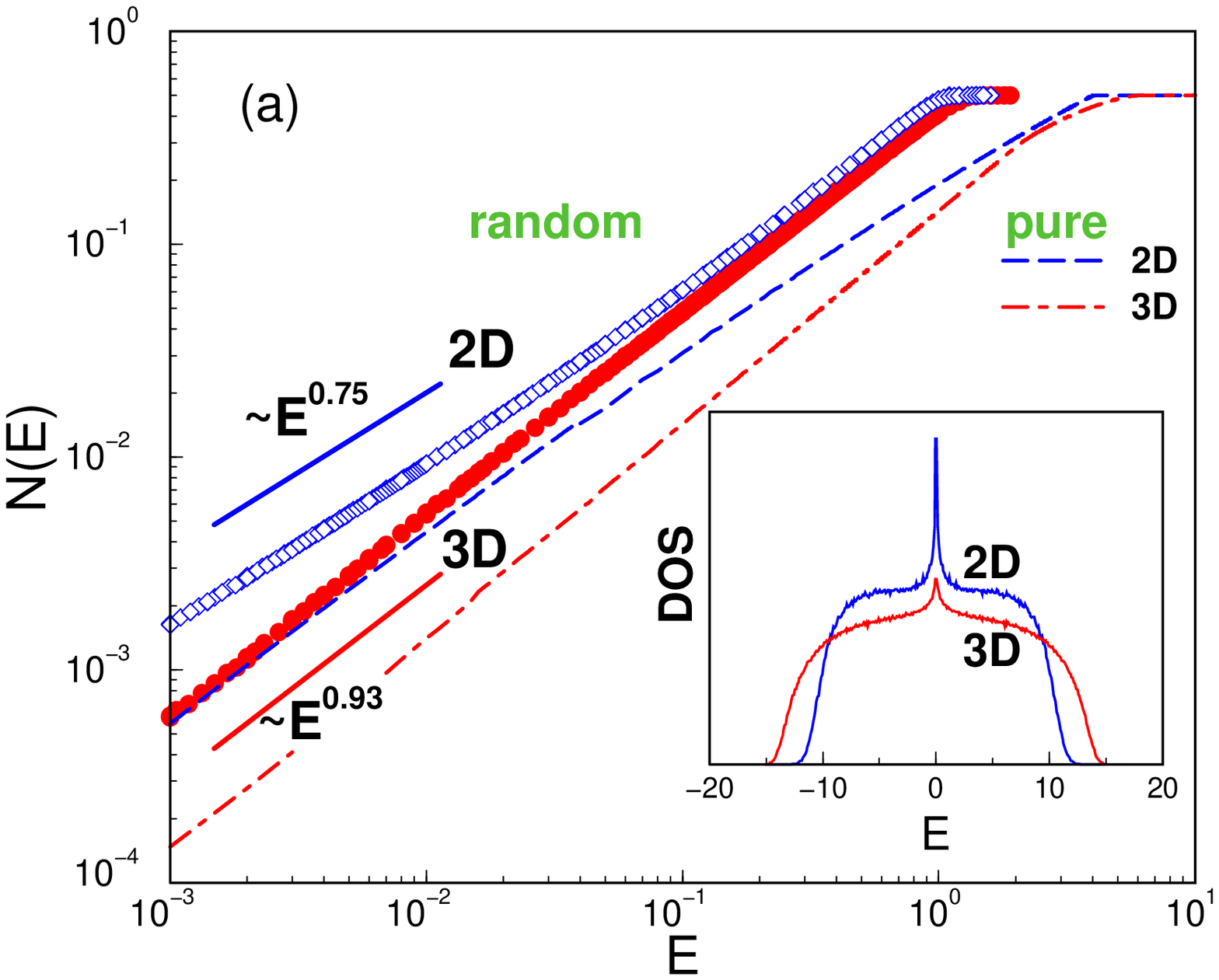,width=8.0cm}}
\centerline{\psfig{figure=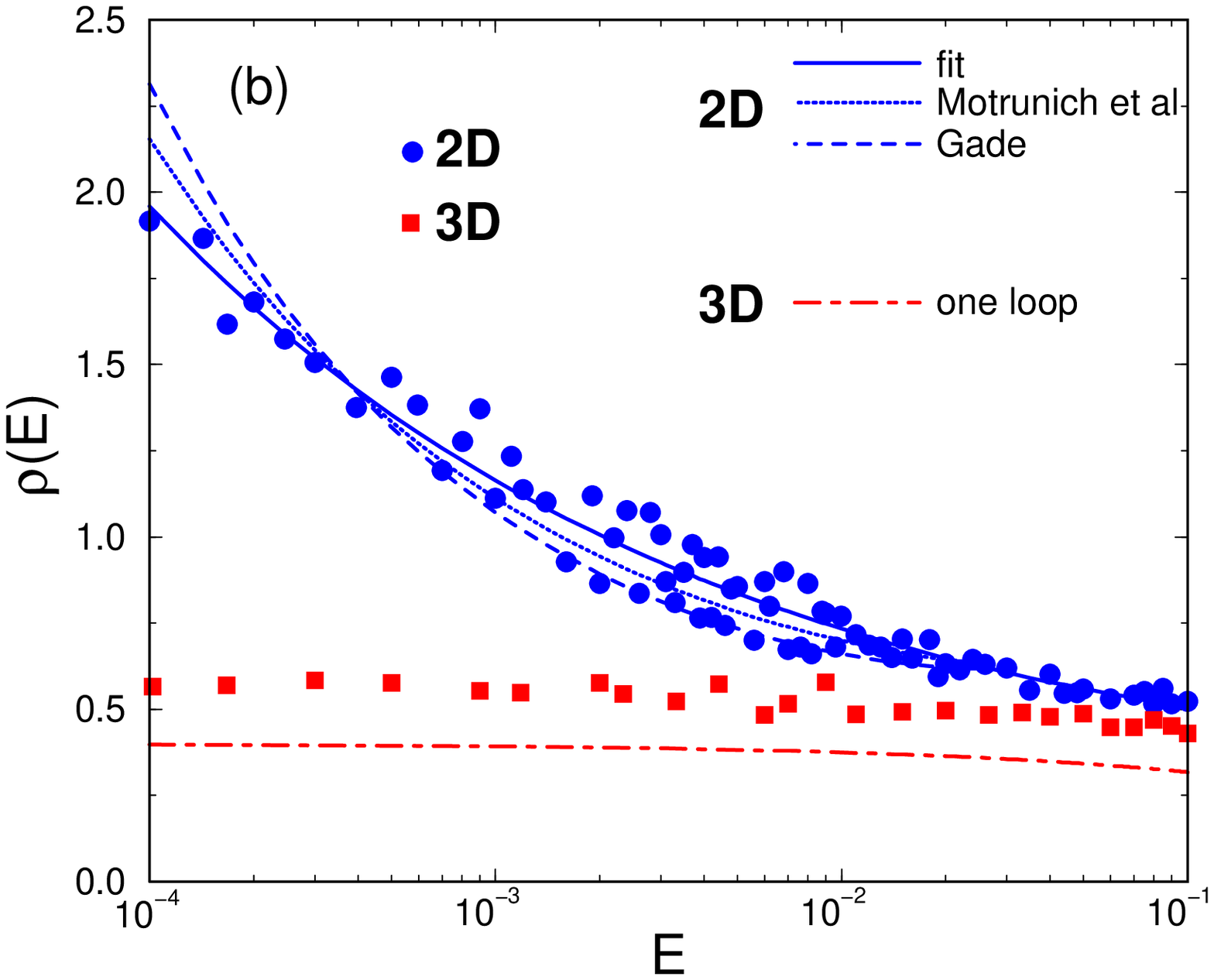,width=8.0cm}}
{\footnotesize{{\bf Fig. 1. (a)}  Log-log plot of the 2D and 3D 
integrated density of states ${\cal N}(E)$, which measures the number 
of eigenvalues from 0 to $E$, with power-law fits of the numerical data
for zero mean  off-diagonal disorder. 
The broken lines give the corresponding idos for the pure systems
while the inset shows linear plots of the densities in the random 
case which display singularities at $E=0$. 
{\bf (b)}  The 2D, 3D density of states $\rho (E)$ 
with the corresponding analytical forms of one loop\cite{32}, 
of Gade \cite{5} with $\kappa=1/2$, of Motrunich et al \cite{24} 
with $\kappa= 2/3$ and our fit with $\kappa\approx 0.91$.
}}

\subsection{The level-statistics}

\medskip
In order to check whether chirality has some effect on the 
localization properties we proceed with the computation of spectral 
fluctuations via the nearest-level $P(S)$ distribution function.
First we ``unfold" the spectrum by making local rescaling of the energy scales
to make the mean level spacing $\Delta(E) \propto 1/\rho(E)$ constant, 
equal to one, throughout the spectrum. This is particularly important
for a singular dos since the mean level spacing tends 
to zero at a singular point. Instead of the raw levels 
$E_{i}$ after ``unfolding" the spectrum consists of the ``unfolded" levels 
\begin{equation}
{\cal {E}}_{i}={\cal {N}}_{av}(E_{i}), \;\; i=1,2,...,N,
\end{equation}
obtained from the averaged integrated spectral density ${\cal {N}}_{av}(E)$. For
the ``unfolded" levels ${\cal {E}}_{i}$ the average nearest-level spacing 
becomes equal to one. Moreover, for the localization properties 
of chiral systems 
we should distinguish between the majority of levels in the spectrum $E\ne 0$ 
and the state at the chiral point $E=0$.

\medskip
In Fig. 2 we present the obtained 2D nearest-level distribution function
$P(S)$ for zero mean off-diagonal disorder,
various sizes $L$ and boundary conditions (BC), using the majority 
of the energy levels in the band except those close to
the band center $E=0$ and the tails. 
In 2D the results for various $L$ seem to lie in-between the Wigner surmise 
(extended states)
and the Poisson distribution (localized states), described by the intermediate
semi-Poisson distribution $P(S)=4S\exp(-2S)$ \cite{15}. For small spacing
$P(S)$ is Wigner-like and for large spacing 
the tails seen in the figure insets are Poisson-like. 
We have also performed averages of $P(S)$ over BC 
which can be also seen to be roughly described by 
the semi-Poisson curve \cite{15}. 
It must be stressed that this distribution appears only for intermediate 
sizes  with $L$ smaller or close to the localization length 
$\xi$ \cite{16,31,35} in the considered energy window.
The obtained results also agree with Ref. \cite{36}
where it is shown that the majority 
of the states for finite 2D disordered systems are multifractal
for certain sizes $L$ below $\xi$ ($L\leq \xi$). 

\medskip 
%
\centerline{\psfig{figure=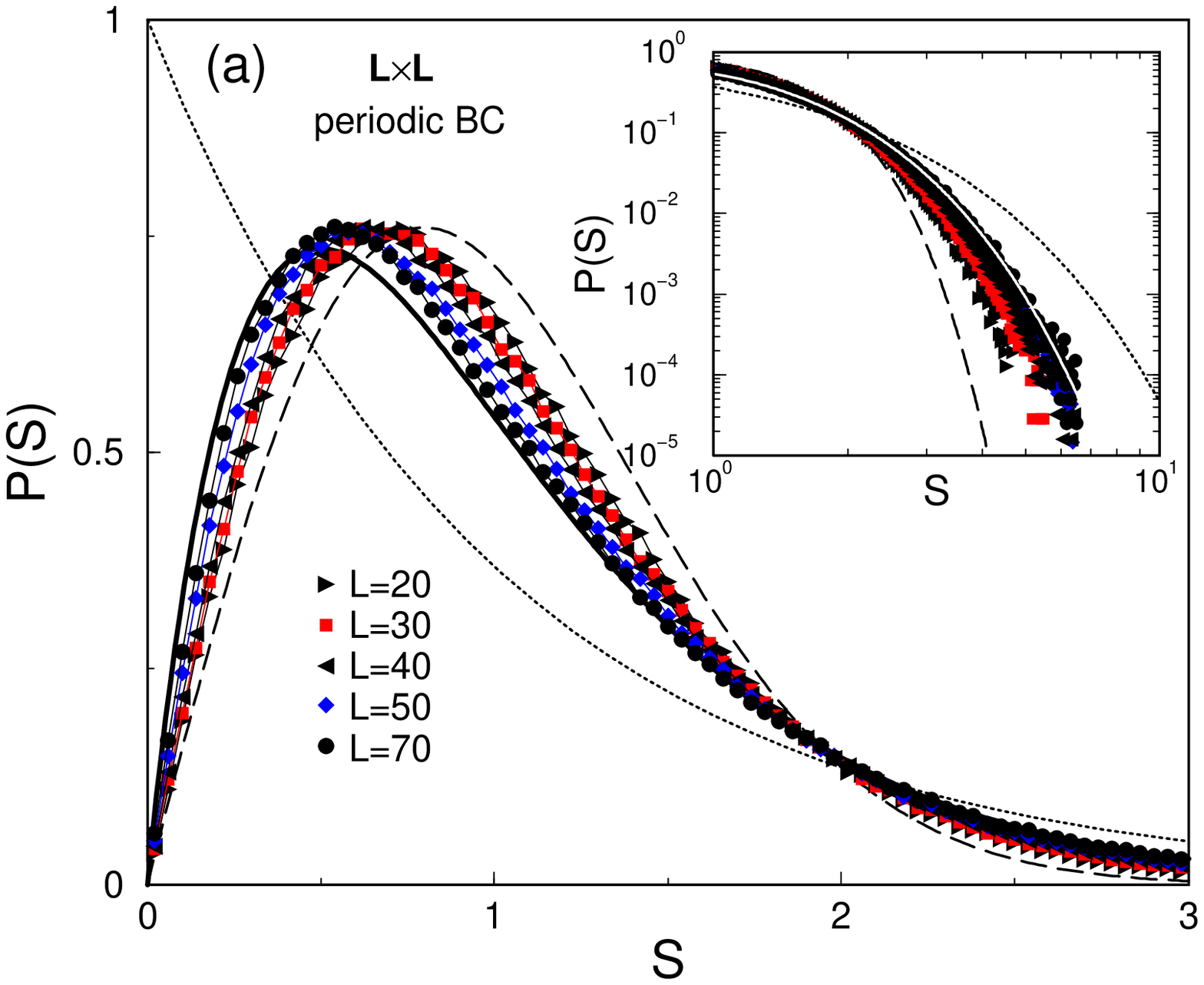,width=8.0cm}}
\centerline{\psfig{figure=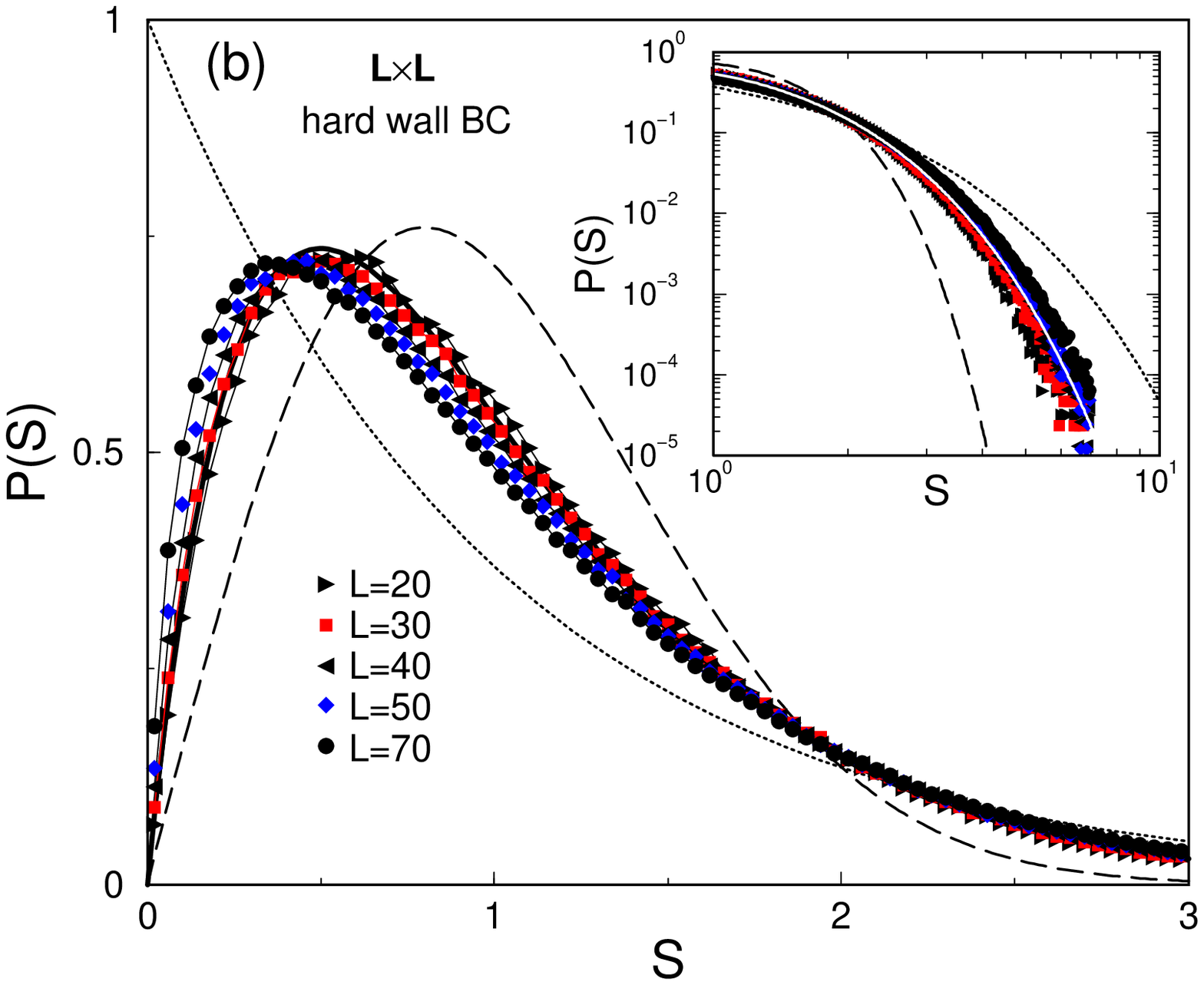,width=8.0cm}}
\centerline{\psfig{figure=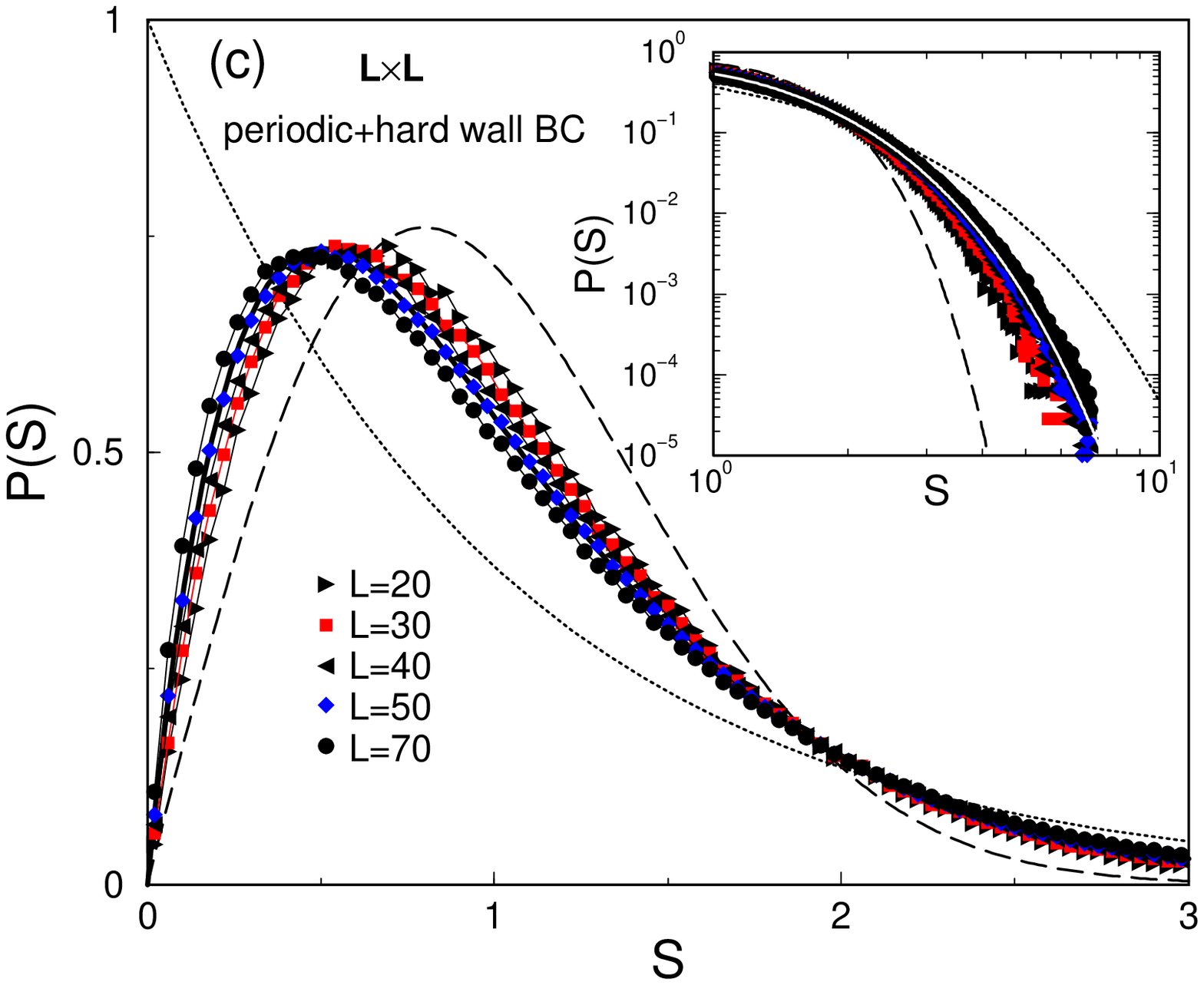,width=8.0cm}}
\centerline{\psfig{figure=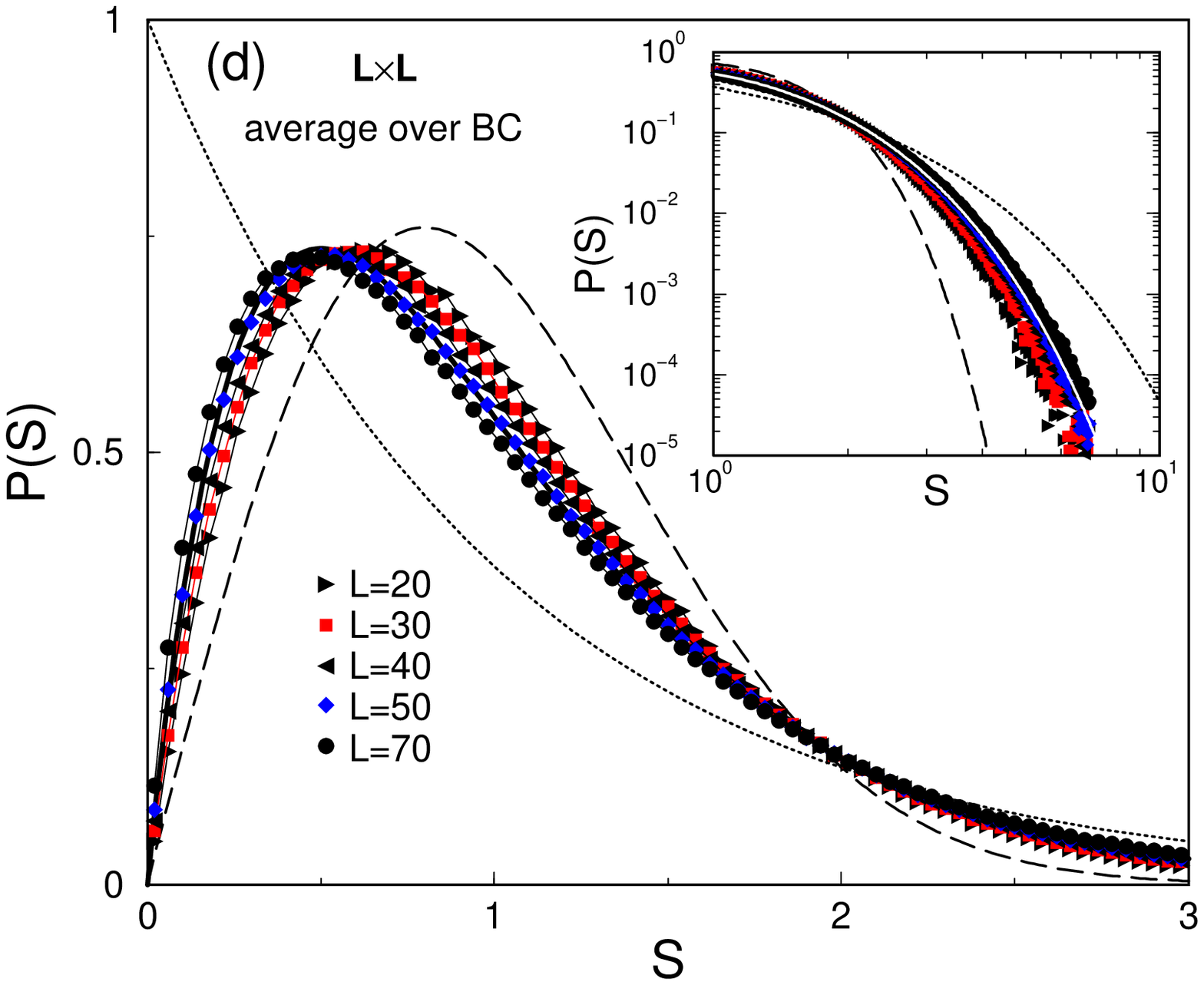,width=8.0cm}}
{\footnotesize{{\bf Fig. 2(a),(b),(c),(d).} The 2D nearest 
level-spacing distribution function $P(S)$ for zero mean 
off-diagonal disorder, various sizes $L$ 
and three possible  boundary conditions, such as periodic (PBC), 
hard wall (HWBC), periodic+hard wall and also average over them. 
The dashed line is 
the Wigner surmise $P(S)=(\pi/2)S \exp(-(\pi/4)S^{2})$, the dotted line is 
the Poisson law $P(S)=\exp(-S)$ and the solid line is the intermediate
critical semi-Poisson distribution $P(S)=4S \exp(-2S)$. 
In the insets we display log-log plots for
the corresponding tails of $P(S)$. For $L=20, 30, 40, 50, 70$ 
the number of runs is $5000, 6000, 10000, 3555$, the
eigenvalues were located in the energy band between $[2,6]$,
for $w=10$ where the full energy range is $[0,13]$,
and their total number was a few million.

}}

\medskip
In order to demonstrate the effect of stronger
level-repulsion close to the chiral $E=0$ state 
we choose as an example a 1D chiral system with off-diagonal disorder
for $L>\xi$ and odd number of sites for the $E=0$ state to exist.
In Fig. 3 we plot the corresponding 1D spacing distribution $P(S)$
of the spacings $S$ defined from the differences between 
the first from the zeroth ($E_{1}-0$), 
the second from the first ($E_{2}-E_{1}$) 
and the eleventh from the tenth ($E_{11}-E_{10}$) 
eigenvalues for logarithmic disorder of $W=1$. This involves 
``unfolding" over the random statistical ensemble since the 
statistics of each spacing $S$ is studied for many random 
configurations at fixed energy. We have additionally performed 
``unfolding" over energy by using the well-known analytical 
form ${\cal {N}}_{av}(E)=(W^{2}/9)/(ln |E|)^{2}$ in 1D
\cite{8}, working instead with the ``unfolded" levels 
${\cal {E}}_{i} \propto 1/(ln |E_{i}|)^{2}$, $\;\; 
i=1,2,...,N$. We observe that the spacing distribution 
of the first positive eigenvalue from the zeroth shown in (a)
is distinguished from the other spacing distributions (b), (c)
and clearly shows a higher degree of level-repulsion. 
In this case the localization length $\xi$ 
can be also obtained for every energy $E$ 
from the expression $\xi(E)=(12/W^{2})$ $|ln |E||$ (e.g. see \cite{8}). 
In Fig. 3 the considered size $L$ is always larger than 
$\xi$ but  $P(S)$ which depends only on the ratio $L/\xi$ 
reaches the Poisson distribution only asymptotically, for $L$ 
much larger than $\xi$. We emphasise again that for  $E$ 
close to the chiral energy $E=0$ (where $\xi$ is larger) 
the Poisson law for 1D localization is seen to be reached 
more slowly.

\medskip 
%
\centerline{\psfig{figure=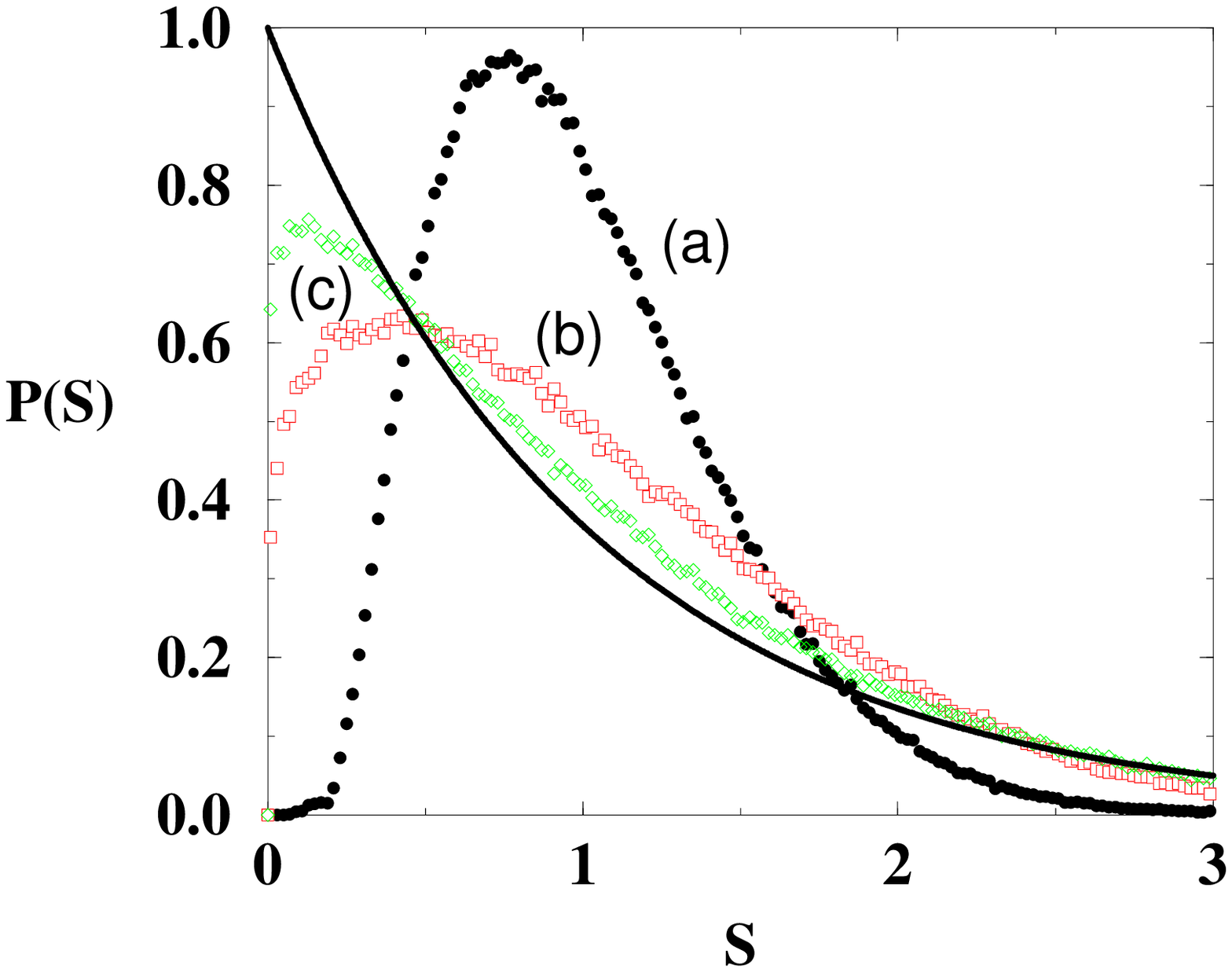,width=8.0cm}}
{\footnotesize{{\bf Fig. 3.} The 1D nearest 
level-spacing distribution function $P(S)$ of (a) the first from
the zeroth (with localization length $\xi \approx 105$), 
(b) the second from the first 
($\xi \approx 75$) and (c) the eleventh from the tenth 
($\xi \approx 36$)
eigenvalues for logarithmic off-diagonal disorder $W=1$ 
and larger size $L=1001$.  The continuous line is 
the asymptotic Poisson law $P(S)=\exp(-S)$ valid for localized
states in the limit $L\gg \xi$ and the total number of runs 
was $500000$.
Note the double ``unfolding" performed
over both disorder and energy since we are very close to 
the singular dos.

}}

\medskip
In Fig. 4(a),(b) we return to the 2D zero mean case and 
plot the probability amplitude distribution of two 
typical multifractal wavefunctions at energies $E\ne 0$. 
They are seen to spread over space with  fractal characteristics 
described by the fractal dimension $D_2$ and
display many scattered peaks (scars). 
It must be pointed out for $E\ne 0$ the states are multifractal 
only for $L<\xi$ where $\xi$ is their localization length, 
while the states at $E=0$ where $\xi$ diverges are described 
by multifractal dimensions valid for any size $L$ \cite{35}.
In order to determine $\xi$ instead of the exponential decay  
of the wavefunction amplitude on very long strips, 
alternatively we can count the number of sites where significant
amplitude exists via the inverse participation ratio (IPR)\cite{13,35}. 
The computed IPR is a rough measure of $\xi$ since it 
favors only the large amplitudes rather than the exponential tails
of the amplitude distributions.
The results for the average IPR in
2D are displayed vs. energy $E$ in Fig. 4(c), (d) for various 
sizes $L$ with hard wall and periodic BC, respectively.
They allow to estimate that localization is not yet reached 
for the studied energy region with the considered sizes $L$ 
lower or equal to the localization length $\xi$. 
For example, in Fig. 4(c),(d) we can see 
that for the largest size $L=100$ the mean IPR is close to
converge, which implies that $\xi$ is roughly 
of the same order of magnitude ($L\approx \xi$).
In Fig. 4(c),(d) localization should be reached only
when the curves of the mean IPR for two different sizes coincide.
We can also speculate that the semi-Poisson appears 
because in the considered energy range for  $P(S)$ 
in Fig. 2 we encounter a  crossover situation 
where $\xi$ which does not vary a lot with $E$
obeys $L\lesssim \xi$, that is $L$ roughly smaller or equal to $\xi$.
The intermediate semi-Poisson distribution found in 2D 
apart from intimately connected to the underlying multifractality
of the wavefunctions (for $L<\xi$) can be regarded as 
a crossover distribution with $L \approx \xi$.
We emphasize that the adopted energy range lies in the middle
between $E=0$ and the tail with the chosen sizes 
$L$ either smaller or close to the wavefunction 
radius estimated by the corresponding $\xi$.

\medskip
%
%
\centerline{\psfig{figure=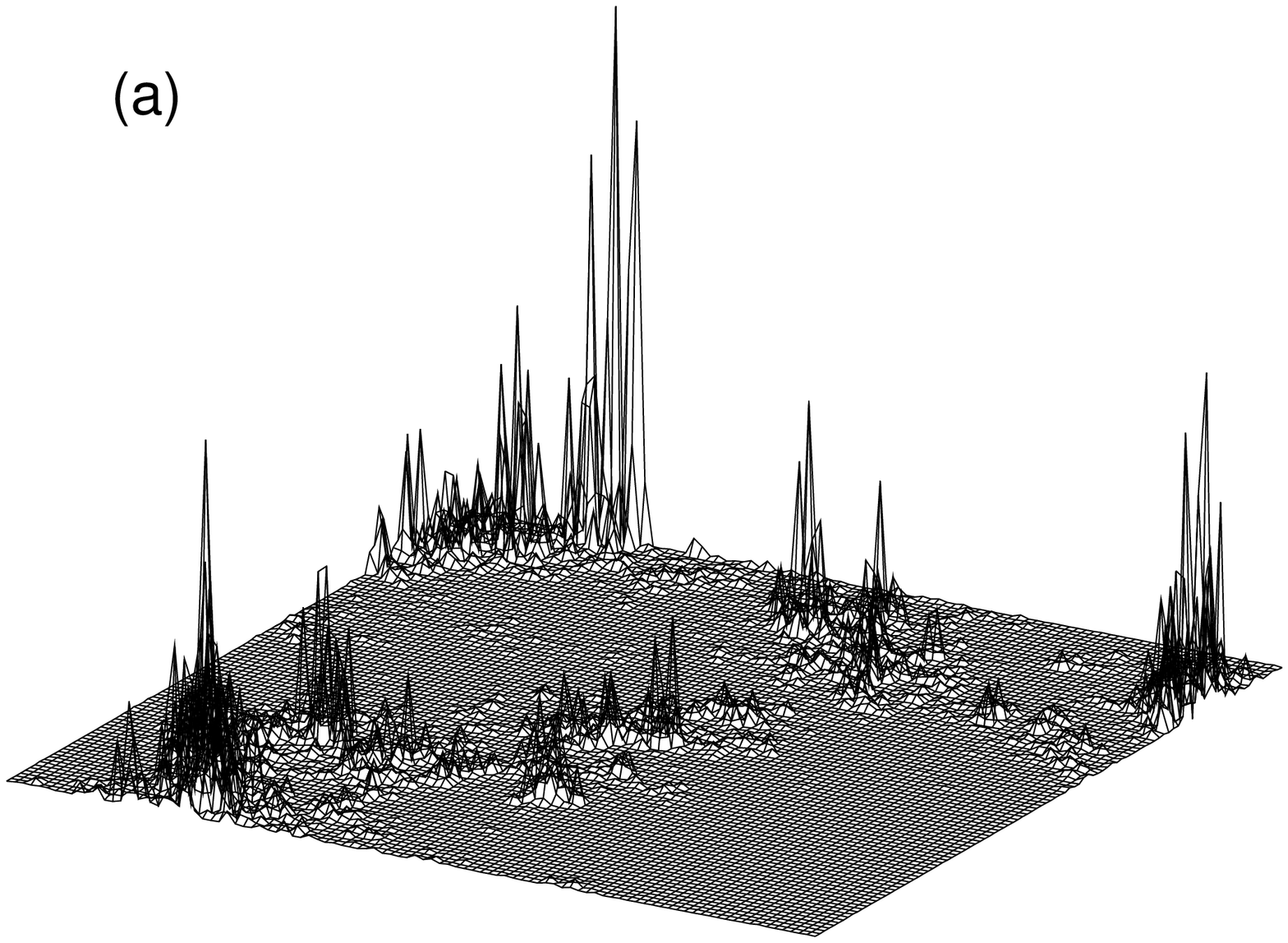,width=8.0cm,angle=-90}}
\centerline{\psfig{figure=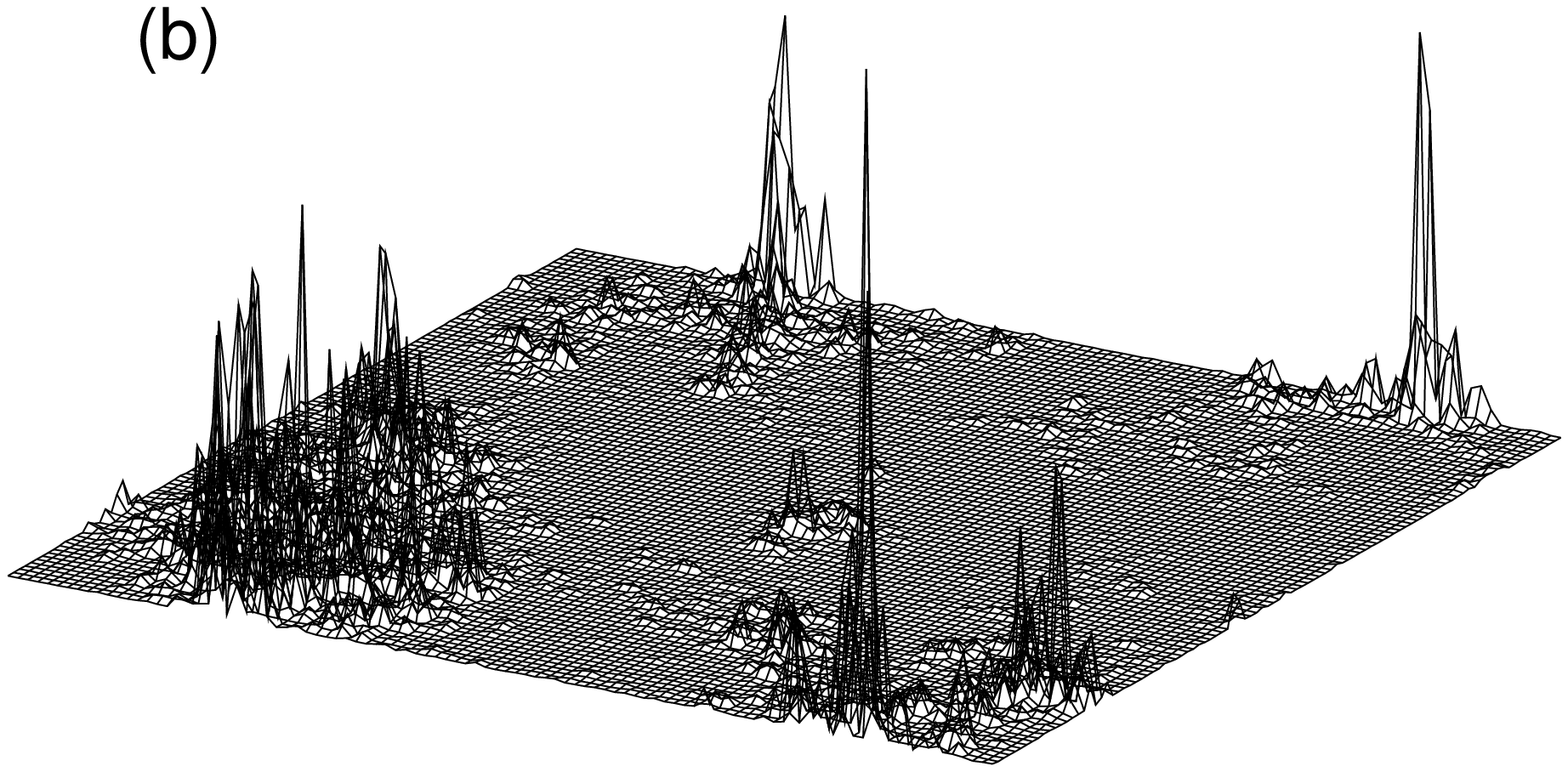,width=8.0cm,angle=-90}}
\centerline{\psfig{figure=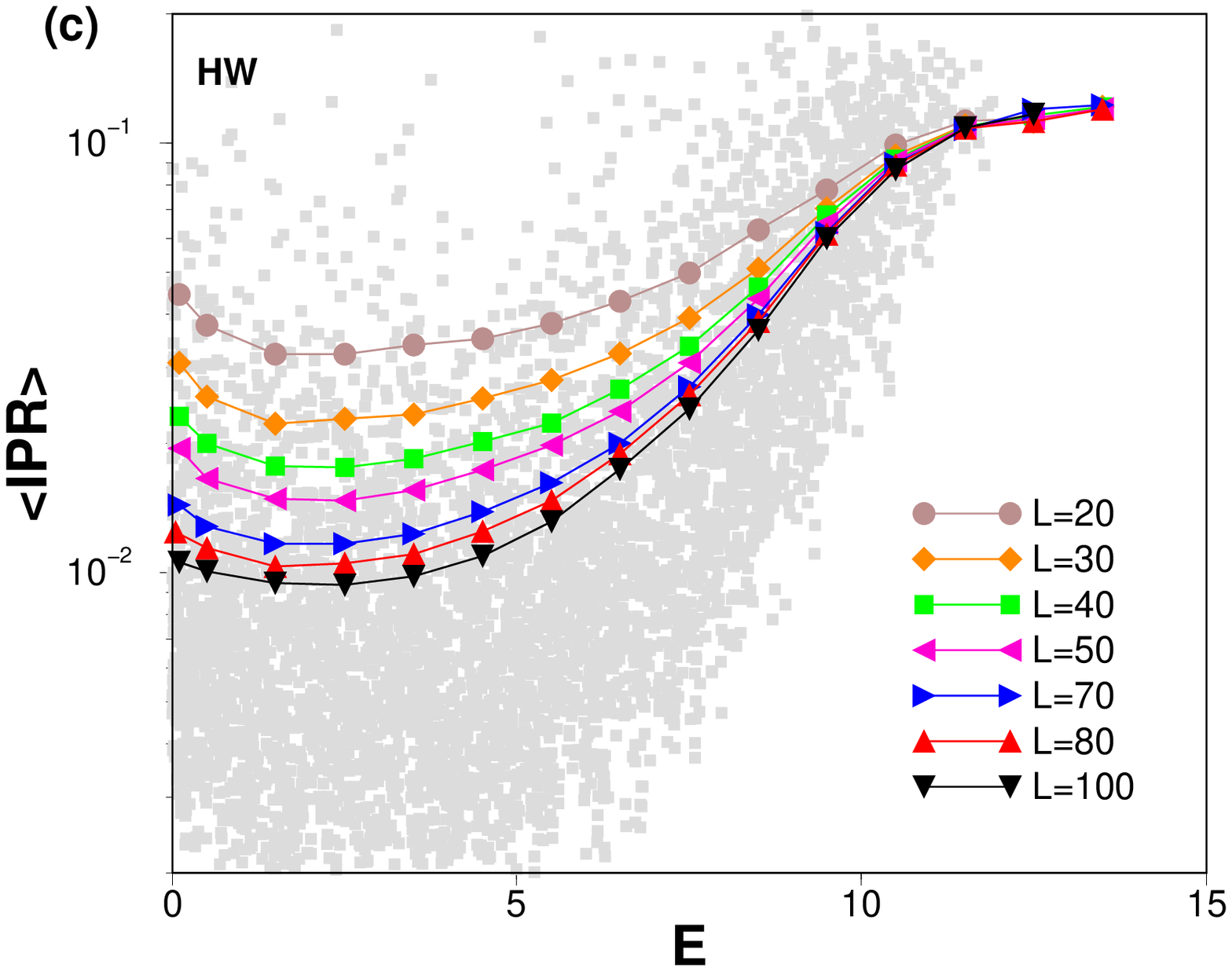,width=8.0cm}}
\centerline{\psfig{figure=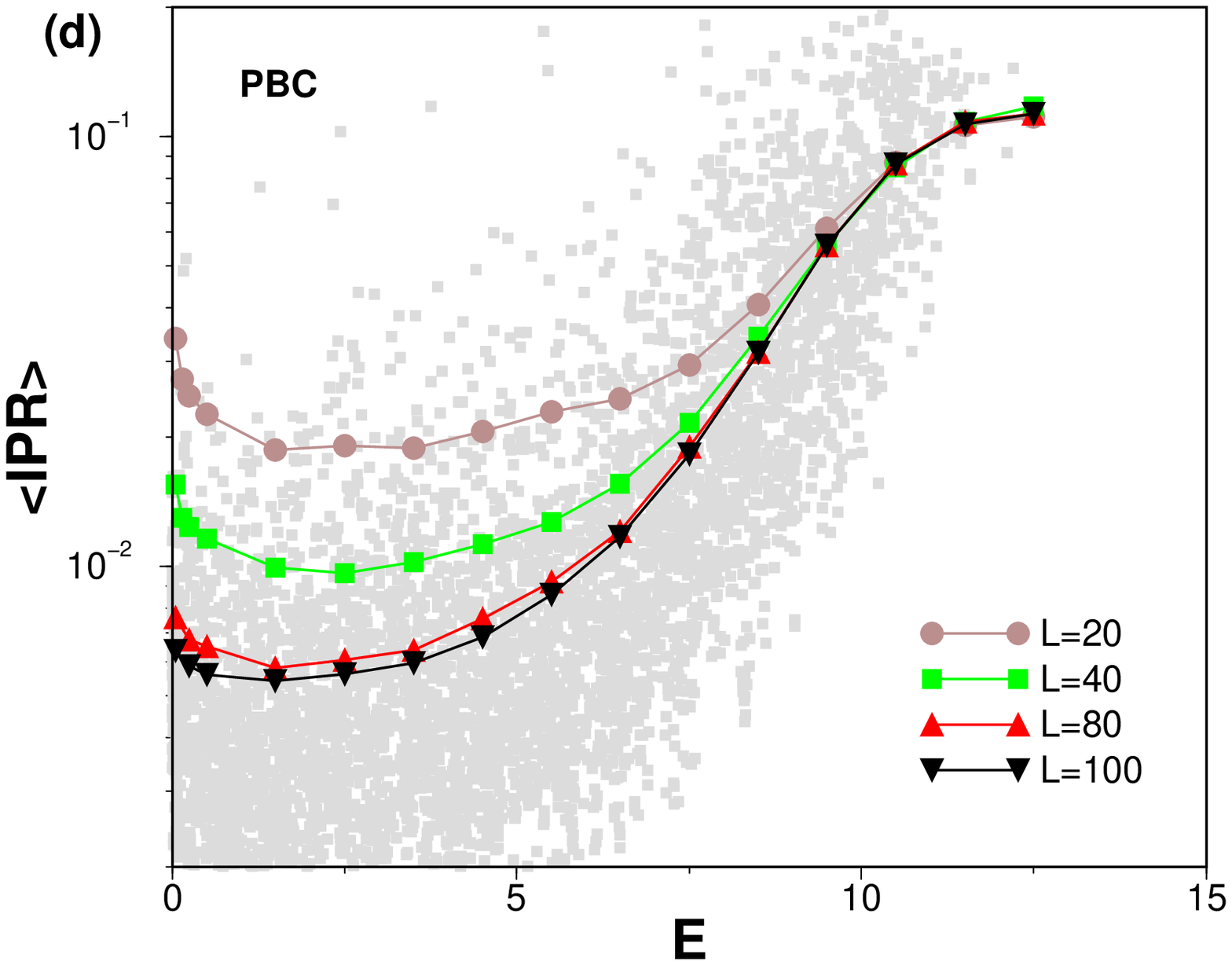,width=8.0cm}}
{\footnotesize{{\bf Fig. 4(a),(b).} The  amplitude distributions of two
typical eigenstates for zero mean off-diagonal disorder
in 2D away from the band center $E\ne 0$. 
The states are multifractal only on scales below the localization $L<\xi$ 
length \cite{36}. The behavior of the $E=0$ state when it exists is 
different since this state is critical for any size with fractal dimension $D_{2}$ 
ranging from the space dimension $D_{2}=d$ for weak off-diagonal disorder 
to $D_{2}=0$ for strong off-diagonal disorder \cite{13}.
{\bf (c),(d)}.  The corresponding average inverse participation 
ratio $\langle IPR\rangle$ vs. energy $E$ for various chiral
even systems with hard wall and periodic BC, respectively. 
Note the values of $\langle IPR\rangle$ are lower for PBC 
where the states are slightly more extended in comparison. 
For finite systems from IPR $\propto L^{-D_{2}}$ 
near the minimum of the curves we find $D_{2}\approx 0.8$,
slightly higher $D_{2}$ near $E=0$ and much
lower $D_{2}$ (localized states)  
for large $E$. The absence of saturation of the curves implies 
that the localization length is larger than the system
for the adopted sizes ($L\lesssim \xi$). 
In the bacground with grey squares we see the actual values of IPR 
for the size $L=100$ only. They display enormous fluctuations 
and are used for computing the corresponding average
for $L=100$.
}}

\medskip
The obtained semi-Poisson $P(S)$ distribution and the multifractality 
for finite size in 2D concerns  states only within the band ($E\ne 0$) 
for finite $L \lesssim \xi$. For the zero mean off-diagonal disorder 
the chosen sizes $L$ are smaller or equal to the localization length $\xi$.
However, for larger $L \gg \xi$ 
(we cannot reach such $L$ in our 2D calculations)
the data from Fig. 2 should eventually 
approach the Poisson curve which indicates only localized states in 2D.
The very slow approach to the localized Poisson limit for infinite size 
$L\to \infty$ is  expected here except at the band center 
$E=0$ where $\xi$  also diverges \cite{16,35}.
In 3D the obtained $P(S)$ for this zero mean off-diagonal disorder is
shown in Fig. 5 to lie very close to the Wigner surmise, a result 
compatible with the existence of mostly delocalized states in 3D.
This implies the asymptotic approach towards Wigner for this
kind of off-diagonal disorder.
In other words, the obtained behavior exhibited by $P(S)$ for $E\ne 0$ 
suggests critical  behavior only for finite systems in 2D 
and delocalized diffusive behavior for any size in 3D. 
Therefore, the localization properties of chiral
systems outside the chiral energy $E=0$ in any dimension 
is similar to that of ordinary disordered systems with the
corresponding localization length $\xi$.

\medskip 
%
%
\centerline{\psfig{figure=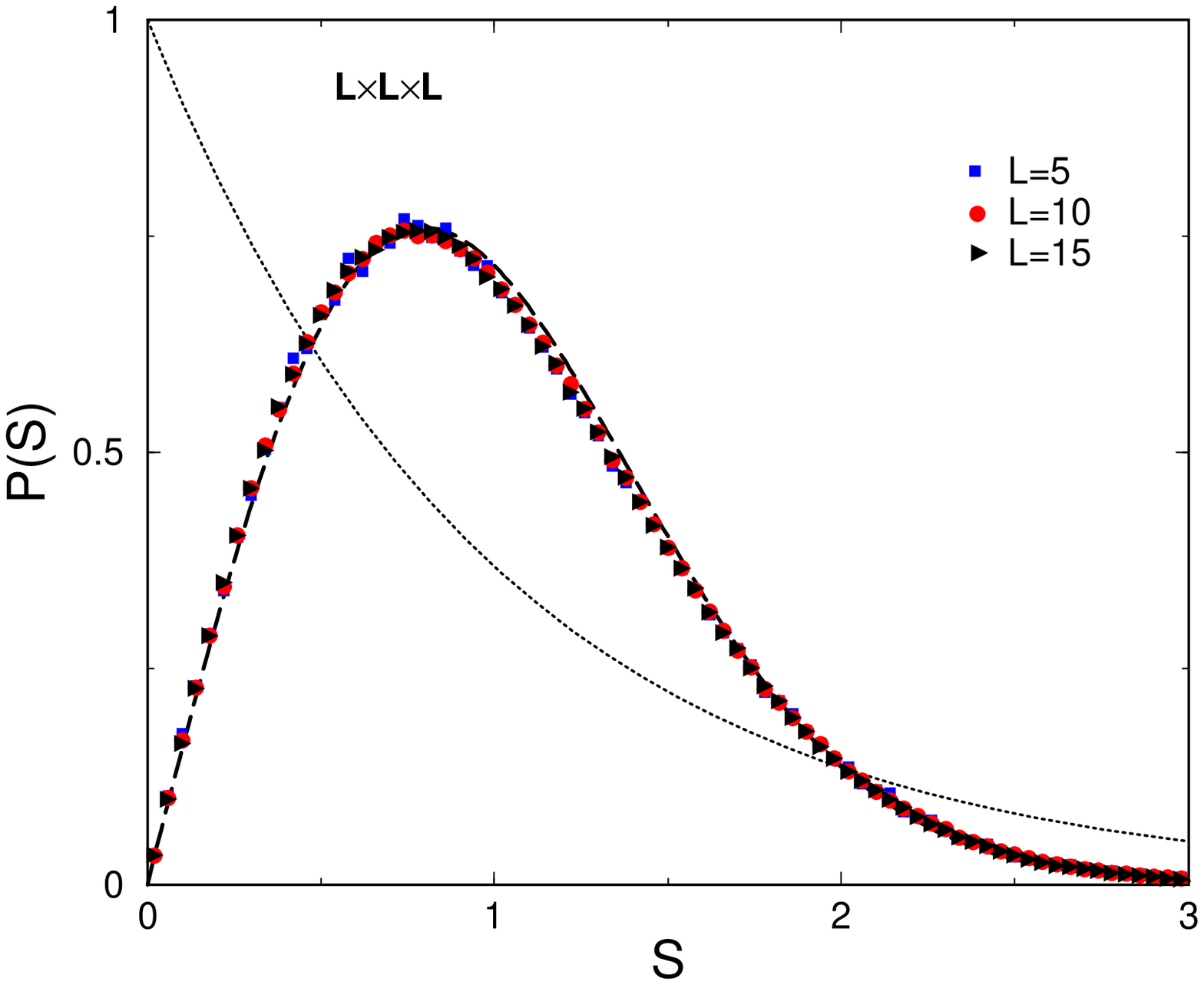,width=8.0cm}}
{\footnotesize{{\bf Fig. 5.} The 3D nearest level-spacing distribution
function $P(S)$ for the zero mean off-diagonal disorder,
various system sizes $L$ with the dashed line giving
the Wigner surmise $P(S)=(\pi/2)S \exp(-(\pi/4)S^{2})$. The sizes 
$L=5, 10, 15$ shown are diagonalized for $5000, 6000, 2000$ configurations
and a few million total number of  eigenvalues was considered
in each case.
}}

\medskip
We have also checked the probability distributions for the first 
few eigenvalues close to $E=0$ 
by employing many random configurations (realizations of disorder) 
as done in \cite{35}. This demonstrates nicely the level-repulsion 
(not shown) by using data for lattices of even number of sites
with periodic BC and odd-number of sites with hard wall BC, 
which preserve chiral symmetry. Although chirality 
is present in both cases the $E=0$ state exists 
only for the later odd case. We  have observed 
the increase of the eigenvalue density $\rho(E)$ close to $E=0$ 
for both even and odd sizes but could not draw any 
conclusions for the statistical behavior of the
energy levels since energy level-repulsion
can be examined only after ``unfolding". 
For the even size $L$ systems the two eigenvalues closest to $E=0$ 
(the lowest positive one and the one with exactly the same energy 
but minus sign) always show level-repulsion due to chiral symmetry. 
The level-repulsion between the two energies $E_{1}, -E_{1}$ 
for the smallest possible $E_{1}$ is expected since these energies 
are always separated by the chiral symmetry point $E=0$. 
On the other hand, the effect of level-repulsion should also be
enhanced for states close to $E=0$ where
the localization length $\xi$ diverges (longer $\xi$
implies stronger level-repulsion). For odd size $L$ 
with chiral symmetry 
the $E=0$ eigenvalue exists for every  random configuration.
The conclusion which can be drawn for this chapter 
is that scaling of level-statistics for $E\ne 0$ chiral systems 
is determined by the ratio of the system 
size $L$ over the corresponding localization length $\xi$, 
similarly to what happens for ordinary disordered systems
and a critical state exists  at $E=0$.

\section{LOGARITHMIC OFF-DIAGONAL DISORDER}

\subsection{The $1/|E|$ leading singularity}

\medskip
In 2D the states away from the band center ($E\ne 0$) are 
also weakly multifractal on scales below the localization length.
The discontinuous behavior claimed for 
the fractal dimension $D_{2}$ at the band center in \cite{35}
is probably due to the different boundary conditions used for the $E=0$ 
and $E\ne 0$ states. At $E=0$ where the localization length 
diverges the corresponding
state is multifractal for every length scale with a fractal 
dimension $D_2$ dependent on the strength of  the logarithmic 
off-diagonal disorder $W$, ranging from $2$ to $0$ for very weak 
and very strong off-diagonal disorder, respectively. 
For strong off-diagonal disorder of width $W$ for the $\ln t_{i,j}$
the observed accumulation of levels close to $E=0$ behaves differently. 
The data shown in Fig. 6 for 2D, 3D can be fitted to power-law singularities  
for the idos 
\begin{equation}
{\cal {N}}(E)\propto |E|^{1-\phi} \Leftrightarrow \rho(E)\propto |E|^{-\phi},  
\end{equation}
with the exponent equal to $\phi$. The ${\cal {N}}(E)$ and the fitted exponents 
$\phi$ are shown in Fig. 6 where the $1/|E|$ Dyson singularity which implies 
$\phi=1$ is approached when $W$ increases. This indicates that the behavior 
depends on the strength $W$ so that for very strongly disordered medium the 
density of states should approach its 1D Dyson limit,
as predicted in \cite{37}.
This is also in agreement with the forms of Gade and Motrunich 
et al forms having very low constant $c_{2}$ in Eq. (4).

\medskip
%
%
\centerline{\psfig{figure=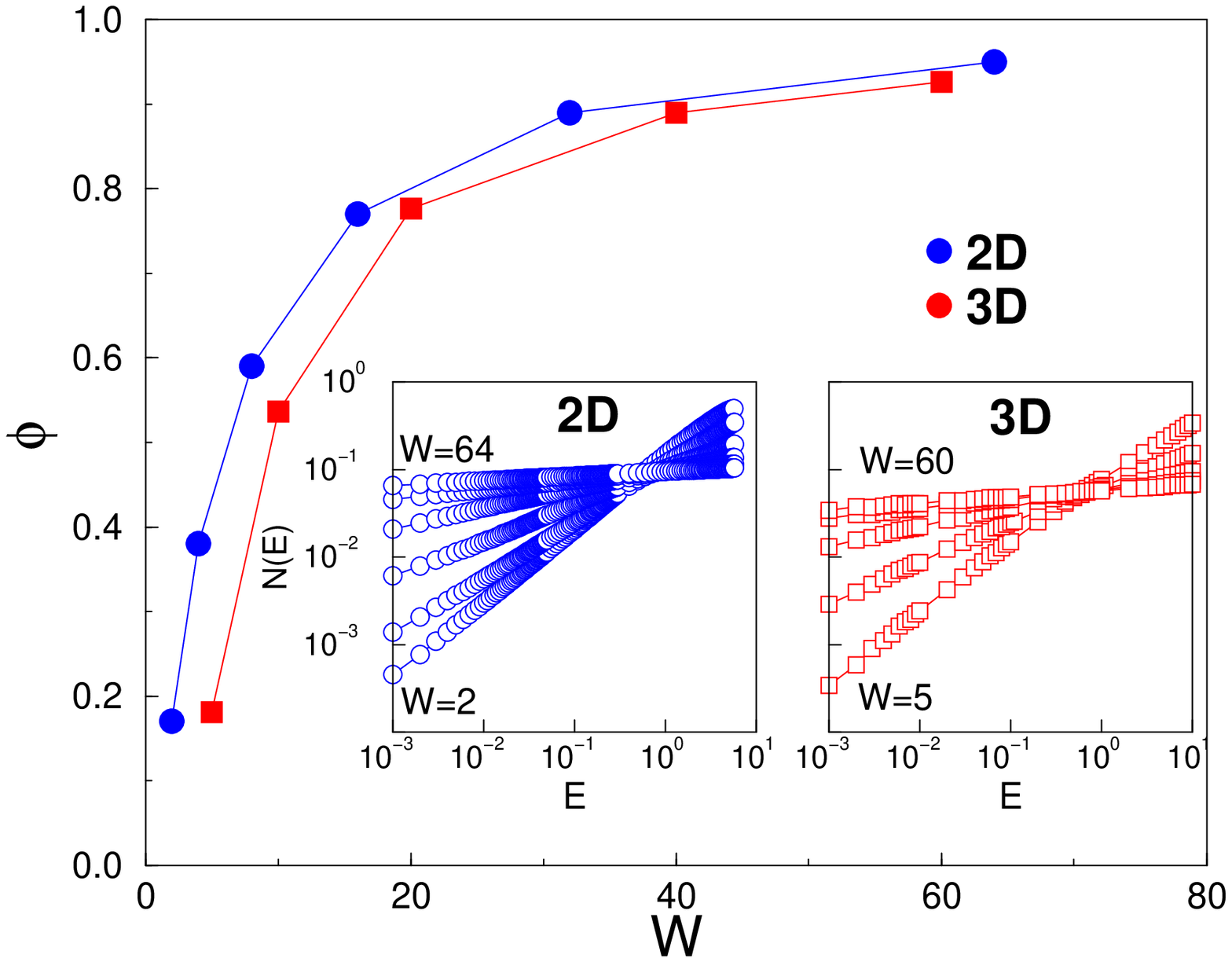,width=8.0cm}}
{\footnotesize{{\bf Fig. 6.} The exponent $\phi$ for the power-law 
fit of the density of state $\rho(E)\propto |E|^{-\phi}$ is shown in 2D, 3D 
to approach the 1D Dyson singularity limit (power-law with $\phi=1$). 
In the inset the corresponding 
idos for various values of the strength of logaritmic
off-diagonal disorder is shown.
Eigenvalue counting algorithms are used for sizes $L=200, 400$ 
and $L=30$ in 2D, 3D,
respectively.
}}

\subsection{The tridiagonalization method}

\medskip
The defined chiral 2D or 3D Hamiltonian $H$ 
can be mapped onto a one-dimensional 
semi-infinite chain via the so-called 
tridiagonalization scheme \cite{31,37}.
The spectral singularities and the degree of localization 
is then reflected on the statistical properties of the created
tridiagonal matrix.
This is a version of the Lanczos 
algorithm \cite{38} which is normally a procedure for finding
eigenvalues and eigenvectors of large sparse matrices 
with most matrix elements zero. The Lanczos method like any other
diagonalization method reduces the matrix into a tridiagonal 
form proceeding as follows: one starts with
a normalized starting state $\Phi_{1}$ 
(e.g. corresponding to a central lattice site) 
and by  operating successively with $H$ generates the 
orthonormalized states $\Phi_{n}$, via the recursion formula
\begin{equation}
\beta_{n} \Phi_{n}=(H-\alpha_{n-1})\Phi_{n-1}-\beta_{n-1}\Phi_{n-2}
\end{equation}
$n=2,...,N$ and $\beta_{1}=0$. The new states $\Phi_{n}, 
n=2,...,N$ are linear combinations of the original lattice basis set,
e.g. in 2D this set consists of the normalized states 
$|n_{1},n_{2}\rangle$, $n_{1}$, $n_{2}$ $=0, 
\pm 1, \pm 2, \pm 3, ...$.
In the new $\Phi_{1}, \Phi_{2}, ...$ basis $H$ is represented  by
a semi-infinite tridiagonal matrix where 
the diagonal matrix elements are $\alpha_{n}, n=1,2,...,N$
and the off-diagonal $\beta_{n}, n=2,3,...,N$. The tridiagonalization
preserves chiral symmetry so that $\alpha_{n}=0, n=1,2,...,N$
for chiral systems. In that case if we start from the central site
corresponding to the normalized state 
$\Phi_{1}=|0,0\rangle$ from Eq. (8) with zero $\alpha_{n}$
we have  $\Phi_{2}=H \Phi_{1}/\beta_{2}$,
$\Phi_{3}=H \Phi_{2}/\beta_{3}- (\beta_{2}/\beta_{3})\Phi_{1}$, etc.
and the $\beta_{n}, n=2,3, ...$ are obtained at each step by normalizing 
the states $\Phi_{2}, \Phi_{3}, ...$.
Since $H$ is a nearest neighbor Hamiltonian
acting on $|n_{1},n_{2}\rangle$ we can arrange the lattice into 
``shells". The state $\Phi_{n+1}$ includes all the lattice sites
up to $n$th  ``shell" (for zero disorder only the sites
of the $n$th ``shell") which in 2D consists of the lattice 
sites $n_{1}, n_{2}=0, \pm 1, \pm 2, \pm 3, ...$ so that
$n=|n_{1}|+|n_{2}|$, starting from $\Phi_{1}= |0,0\rangle$. 
As an example, for the periodic 2D Hamiltonian 
one computes $\beta_{2}=2, \beta_{3}=\sqrt{5},
\beta_{4}=\sqrt{76/20}$, etc, 
approaching $\beta_{n\to \infty}= 2$. 
The basis states for the corresponding tridiagonal matrix are 
$\Phi_{1}=|0,0\rangle$, 
$\Phi_{2}=(|1,0\rangle+|0,1\rangle+|-1,0\rangle+|0,-1\rangle)/\beta_{2}$, 
$\Phi_{3}=(|2,0\rangle+|0,2\rangle+|-2,0\rangle+|0,-2\rangle
+2|1,1\rangle+2|1,-1\rangle+2|-1,1\rangle+2|-1,-1\rangle)
/(\beta_{2}\beta_{3})$, 
etc.

The advantage of the tridiagonalization method is economy of storage since
we can save only the $\alpha$'s and $\beta$'s as we go along. 
The weakness of the method is loss of orthogonality
due to rounding errors (although for sparse matrices 
the damage is smaller). In 2D or 3D if we start from a central site the
matrix consists of blocks around it so that each block
corresponds to a ``shell". A site which can be reached from the 
central site ($0$th ``shell") in a minimum of $n$ steps from the origin
is in the ``shell" $n$.
The 2D or 3D matrix is block-tridiagonal with blocks 
corresponding to ``shells" starting from the $0$th ``shell" 
with $\Phi_{1}$ and we can work with $N$ iterations, 
exactly equal to the number of ``shells",
to determine $\Phi_{2}, \Phi_{3}, ..., \Phi_{N+1}$  
and $\beta_{2}, \beta_{3}, ..., \beta_{N+1}$ for the 
$1, 2, ..., N$th ``shells" around the central site. 
From the obtained 1D  semi-infinite tridiagonal matrix 
one can conveniently compute the local density of states on
the central site $\Phi_{1}$ which has
its first $2N$ moments exactly equal to the 
spectral density of the original 2D or 3D matrix.
It must be stressed that the use of the Lanczos
algorithm nowadays is different than that of the
tridiagonalization method. 
Its wide use is directed, instead, towards the computation
of specific eigenvalues of large sparse matrices
with the requested accuracy.
This can be achieved by increasing the number of iterations to 
much larger values than the number of ``shells" or ``blocks"
considered in tridiagonalization.

For zero disorder or disorder which preserves chiral symmetry
connecting $A$ and $B$ sublattices a 2D or 3D tight binding 
system maps to a chain with all $\alpha_{n}=0$. 
For chiral disorder the mapping onto a 
1D chain involves only off-diagonal disorder but inhomogenous
type  ($n-$dependent) \cite{37,31}. It is only for 
very strong off-diagonal disorder where the disordered chain 
resulting from the tridiagonalization becomes
homogenously off-diagonal disordered. On the basis 
of our numerical results
we argue that although this conjecture holds
the  suggested inhomogenous decay in the case of 
not so strong disorder 
is absent as well (or it is not of power-law type).
 
Since for chiral systems the 1D 
semi-infinite chain is also chiral 
with off-diagonal disorder only the well-known 1D 
results could map to higher dimensions. 
The resulting 1D chain from the tridiagonalization
scheme has inhomogenous off-diagonal disorder with 
hopping matrix elements $\beta_{n}$ 
chosen from an site $n$-dependent random distribution 
of mean $\langle \beta_{n}\rangle$ and variance var$(\beta_{n})=
\langle \beta^{2}_{n} \rangle - \langle \beta_{n} \rangle^{2}$. 
Under the approximation that the states $\Phi_{n+1}$
are expressed as linear combinations of 
$|n_{1},n_{2}\rangle$ belonging to the $n$th  ``shell" only,
which is valid for pure systems and claimed to hold for weak 
off-diagonal disorder, 
in \cite{37} a power-law decay of the variance vs. $n$ 
was predicted. This gave $\psi(r)\propto e^{-\gamma \sqrt{\ln r}}$ 
decay of the $E=0$ chiral state in every dimension higher than one.
Our numerics show that some of the arguments which led to this result 
are not valid since for weak off-diagonal disorder the variance vs. 
$n$ does not decay as a power-law, but one eventually obtains 
indications of saturation for large $n$ (although the data 
for smaller $n$ could be fitted to power laws). 
For strong off-diagonal disorder the obtained saturation
implies a constant var$(\beta_{n})$ independent of $n$
and only in this case the results 
for 1D off-diagonal disorder, such as the $1/|E|$ Dyson-type 
spectral singularities, carry through to higher dimensions.

\medskip
%
%
\centerline{\psfig{figure=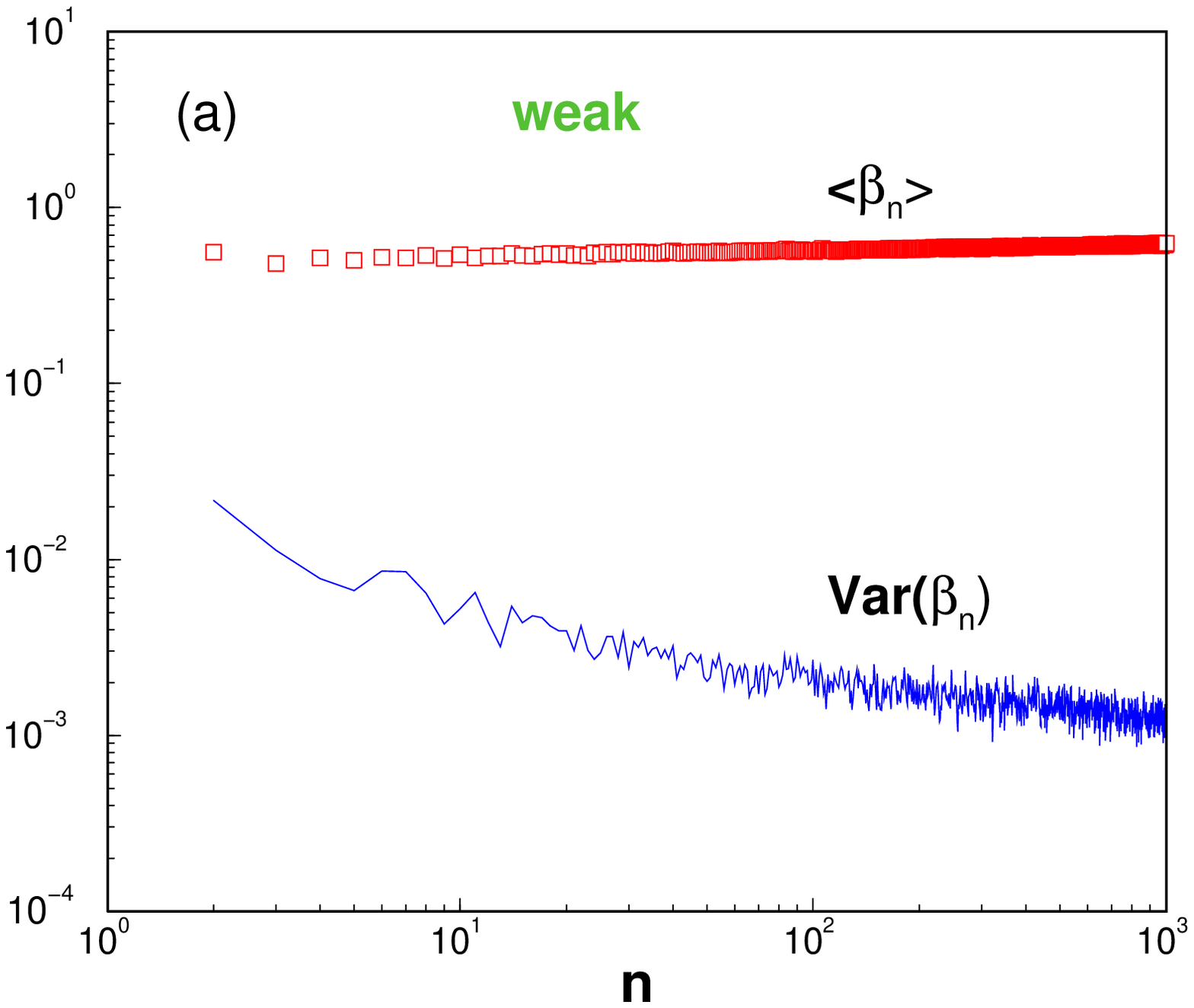,width=8.0cm}}
\centerline{\psfig{figure=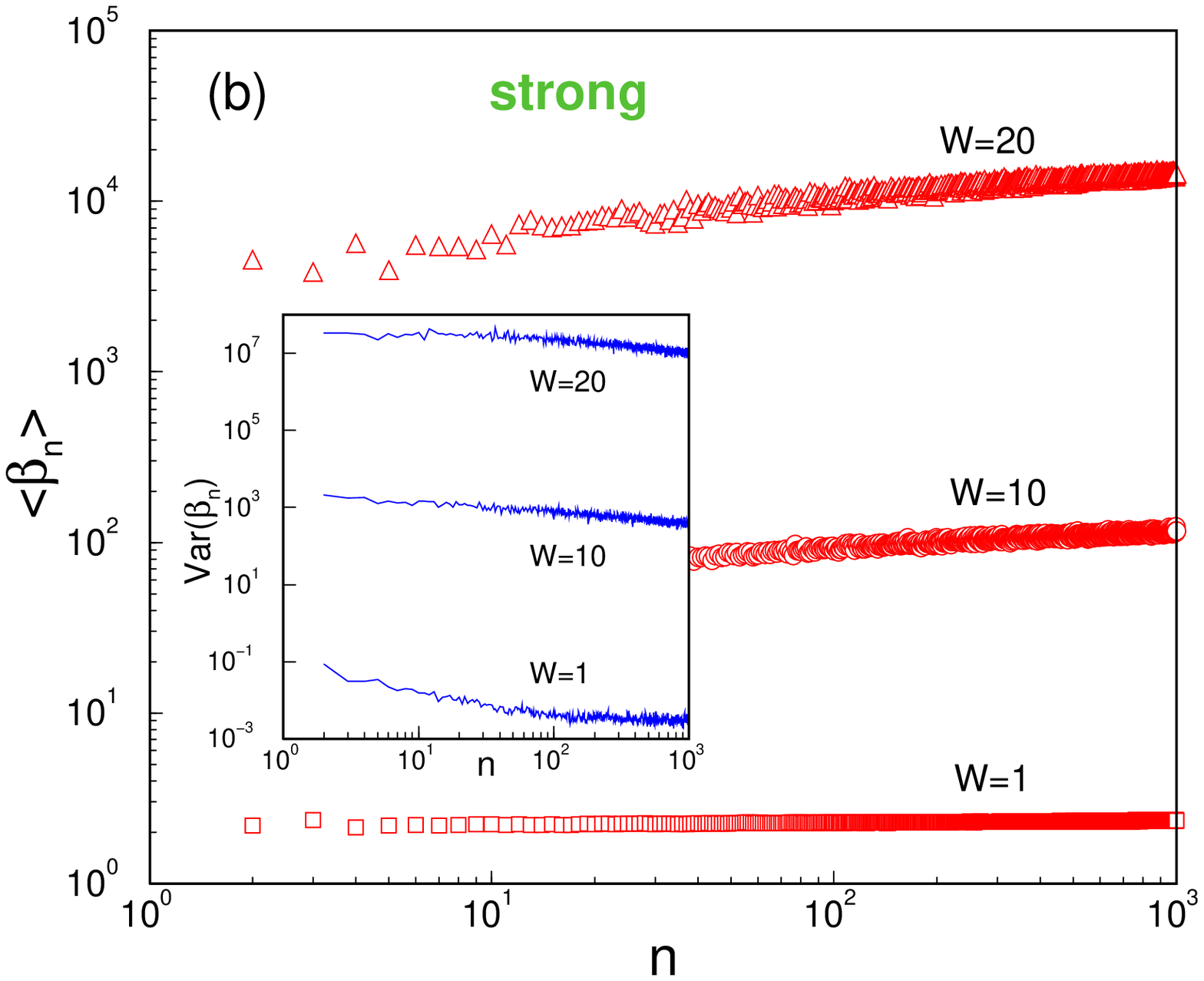,width=8.0cm}}
{\footnotesize{{{\bf Fig. 7(a).} The parameters of the tridiagonal  
chain  with mean $\langle \beta_{n}\rangle$ 
and variance var$(\beta_{n})$ as a function of the chain 
index $n$ for zero mean off-diagonal disorder and $5000$ random
configurations. 
Although decay of the variance is seen for small $n$, 
implying inhomogenous disorder for the chain, 
the data point towards saturation at large $n$. 
{\bf (b).}The parameters of the tridiagonal  
chain $\langle \beta_{n}\rangle$ and their variance 
var$(\beta_{n})$ as a function of the tridiagonal chain index $n$ 
for various values $W$ of logarithmic off-diagonal disorder. 
The saturation of the variance var$(\beta_{n})$  suggests a mapping 
of a higher-dimensional strongly off-diagonal disordered systems
to a random chain. 
}}}

\medskip
In order to check these predictions 
we are able to extend the previous numerical 
calculations to higher number of ``shells" $n$ starting from a
central site $\Phi_{1}$ ($0$th ``shell"). 
For example, for the zero mean disorder in 2D 
we can go up to $N=1000$ ``shells" starting from the site at the origin
and the total number of sites from the starting 
central site ($0$th ``shell") up to the $N$th ``shell" is $2N^{2}+2N+1$.
For  the largest $N$ used we always find saturation of the variance vs. $ n$, 
even for the zero mean off-diagonal disorder (Fig. 7(a)) and certainly 
for the logarithmically strong off-diagonal disorder (Fig. 7(b)). 
The predicted inhomogenous power-law decay of \cite{37}, 
which implies $n$-dependent variance, seems not to hold
but such decay might exist only for finite sizes (small $n$). 
For large enough $n$ for both zero mean and strong logarithmic
off-diagonal disorder the mapping to the 1D chain 
involves $n$-independent (homogenous) disorder. 
The above discussion is not in favor of the square 
root of log exponential law predicted 
in the tridiagonalization of \cite{37} 
and rather points towards the presence of $1/|E|$ forms 
for the dos in higher $d$ predicted in Gade and Motrunich et al.
For strong off-diagonal disorder the $n$-independent variance prediction
with associated 1D behavior in 2D, 3D holds more clearly 
and the $1/|E|$ Dyson divergence should somehow appear. 
The data for the hoppings $\langle \beta_{n}\rangle$ 
and the variance var$(\beta_{n})$ vs.
$ n$ are shown in Fig. 7(a) for zero mean and (b) strong logarithmic
off-diagonal disorder.
Although the saturation of $\langle \beta_{n}\rangle$ can be seen 
in all cases the var$(\beta_{n})$ 
seems to decay for the zero mean off-diagonal disorder.
However, this is true only for small $n$ and the var$(\beta_{n})$ 
is eventually  seen having a tendency to saturate 
for larger $n$ (Fig. 7(a)).
For  large $W$ the saturation already 
occurs for smaller $n$ (Fig. 7(b)). In 3D 
we have obtained similar results.

\section{DYNAMICS OF TWO-INTERACTING PARTICLES WITH OFF-DIAGONAL DISORDER}

\medskip
In Fig. 8(a) we observe the dynamics of one particle for logarithmic
off-diagonal disorder in 1D.
We find that the  effect of off-diagonal disorder turns out to be similar 
to that of diagonal disorder, both deviating from
the pure chain ballistic law $\sigma^{2}(t)=2t^{2}$
(dashed line in Fig. 8(a)) in proportion to the added disorder. 
Next we turn to the interacting case since the chiral symmetry 
(rather its absence) 
was recently shown to be crucial for the
possibility of a metallic phase in 2D. For half-filled interacting systems
the Mott gap turns out to be insensitive to weak disorder when 
chiral symmetry is present, so that a metallic phase in 2D 
could become possible only when the chiral symmetry is broken \cite{29}. 

\medskip
%
\centerline{\psfig{figure=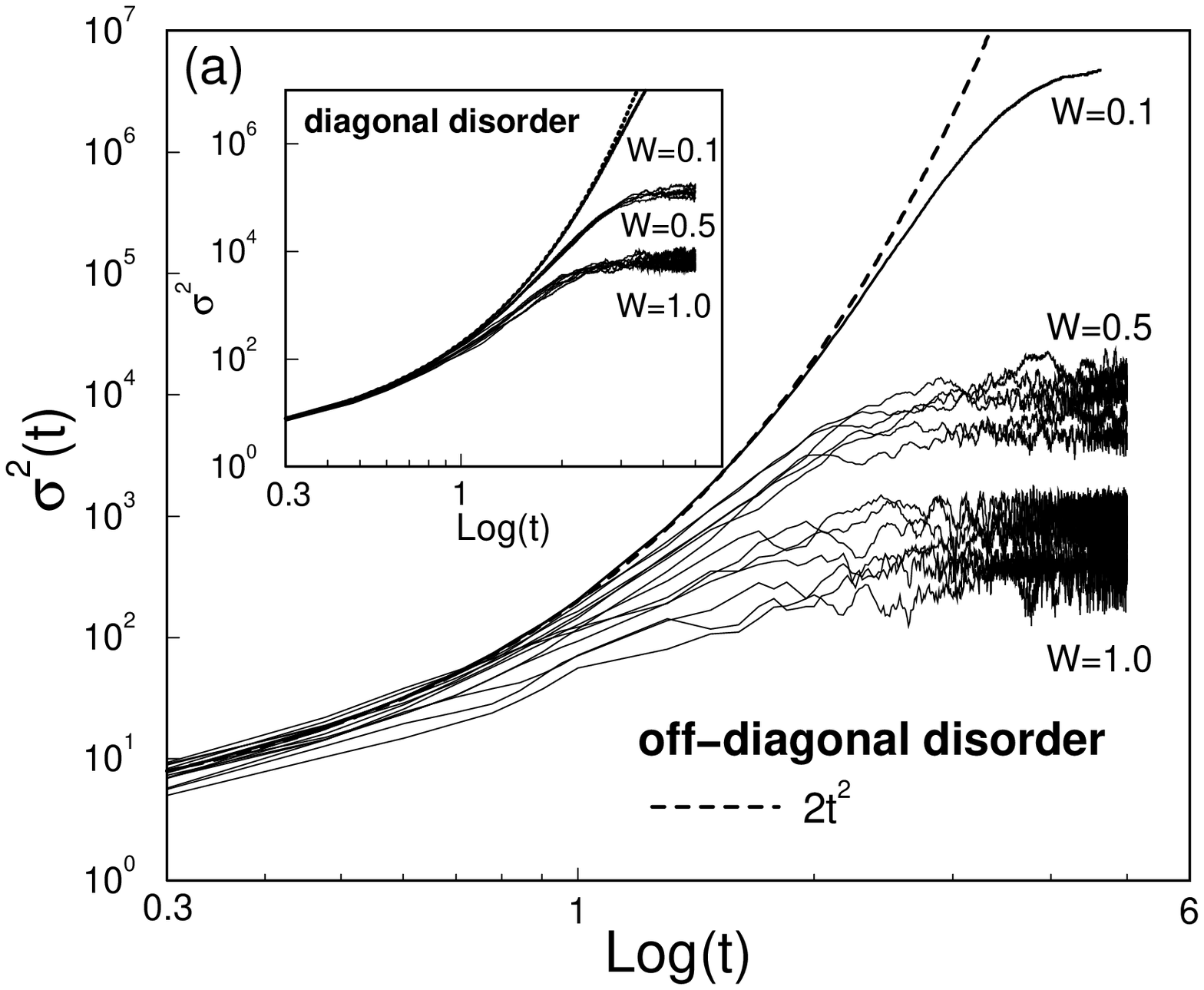,width=8.0cm}}
\centerline{\psfig{figure=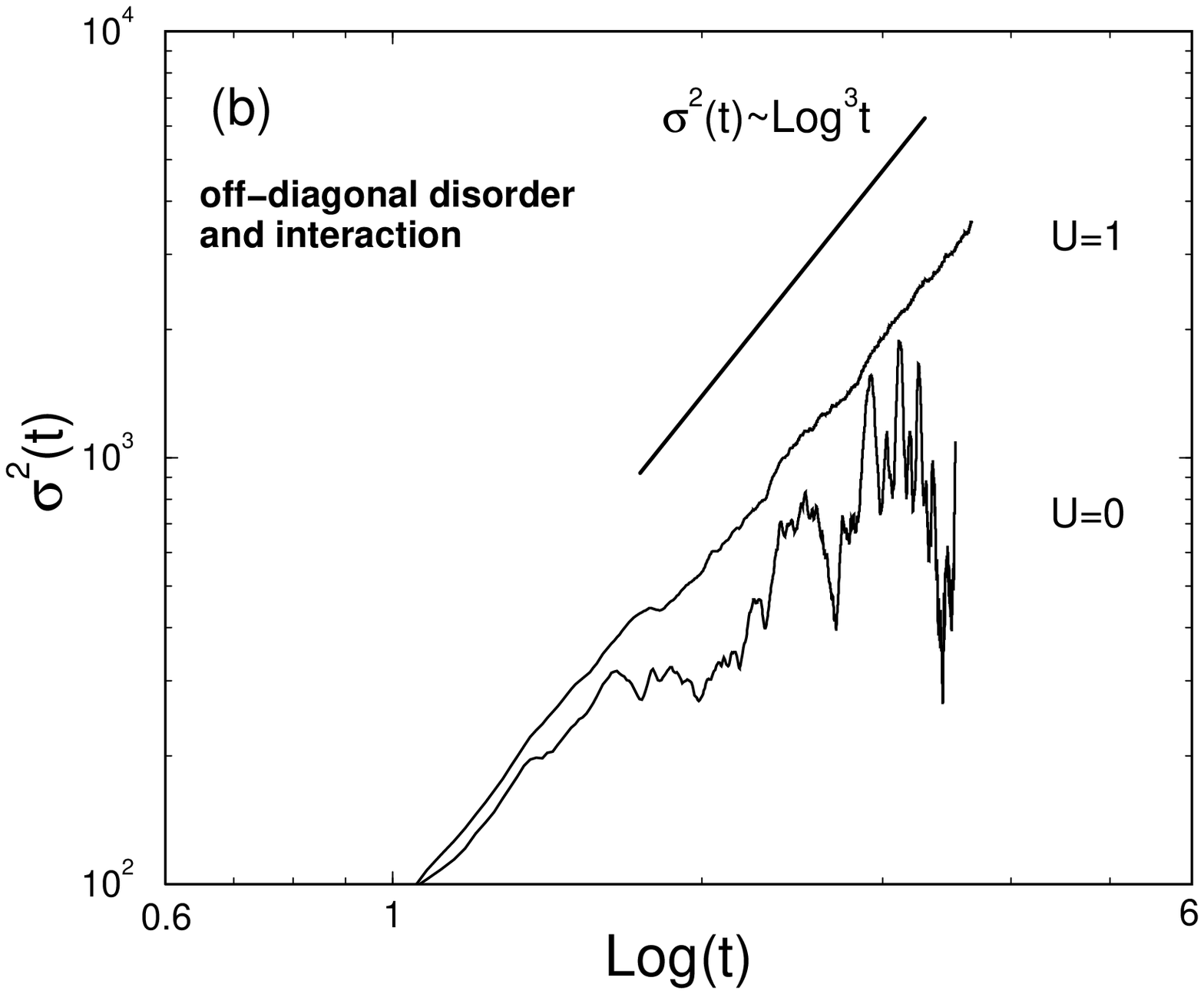,width=8.0cm}}
{\footnotesize{{\bf Fig. 8(a).} Log-log plot of the mean-square displacement 
$\sigma^{2}(t)$ as a function of $\log(t)$ in 1D (note the double
$\log (\log (t))$ in the x-axis so that $1$, $6$ correspond to $t=10$,
$t=10^{6}$, respectively, and the single log $\sigma ^{2}$ in the 
y-axis, all for base ten logarithms) 
for logarithmic off-diagonal disorder with $W$=0.1, $W$=0.5 and $W$=1.
The dashed line is the ballistic limit $2t^{2}$.
In the inset results for diagonal disorder from $[-W/2,+W/2]$ are shown. 
{\bf (b)} A log-log plot of the mean-square displacement $\sigma^{2}(t)$ 
for two particles as a function of $\log(t)$ for logarithmic 
off-diagonal disorder with
$W$=1 and interaction strength $U=0, 1$. The enhancement of the wavepacket 
in the presence of interaction as well as the smoother evolution with
an approximate ultraslow diffusive law
can be seen also for off-diagonal disorder.
}}

\medskip
In order to check the effect of repulsive 
interactions in 1D chain we have computed 
the mean square displacement from the formula
\begin{equation}
\sigma^{2} (t)=(1/2) \sum_{n_{1},n_{2}} 
|\Psi_{n_{1},n_{2}}(t)|^{2} 
(n_{1}-\overline{n}_{1}+n_{2} - \overline{n}_{2})^{2},  
\end{equation}
with $ \Psi_{n_{1},n_{2}}(t)$ the time dependent wavefunction 
for two particles at sites  $n_{1}$, $n_{2}$
in the presence of off-diagonal disorder and interaction $U$
\cite{39,40}, both starting at the same initial zeroth site.
The mean electrons positions are $\overline{n}_{i}= \sum_{n_{1},n_{2}} 
|\Psi_{n_{1},n_{2}}(t)|^{2} n_{i}$, $i=1,2$. 
The result for $\sigma^{2} (t)$ is shown in Fig.
8(b) for off-diagonal disorder without interaction 
($U=0$) and interaction of strength $U=1$. We verify that
the two-particle propagation increases with interaction, 
as for ordinary diagonal disorder, where propagation is known to be
enhanced in the presence of disorder when $U$ is switched on \cite{39,40}.
For increasing disorder we find that propagation decreases 
as expected. In the presence of both disorder and interaction
propagation is also seen to become smooth and the data obey 
an ultraslow diffusion law $\sigma^{2} 
(t)\propto \log^{\alpha}(t)$ with a slope from the linear fit 
equal to $\alpha \approx 3$ (Fig. 8(b)) when $U=1$.
Unfortunately, we cannot discuss the half-filled chiral symmetric limit 
since our case involves only low filling of two particles only.
For $U\ne 0$ our system clearly violates the chiral symmetry
since it is no longer particle-hole symmetric due to $U$ which is added
in the matrix diagonal. 
More generally a bipartite lattice structure with chiral symmetry 
was discussed in the absence of disorder \cite{41} for general filling 
and dimension with interaction strength $U$. 
Their results hold for our chiral systems too since
for negative $U$ the spectrum turns out to be opposite to that for positive $U$
while the energy $E$ corresponds to the energy $-E$ of the other model, 
respectively. 
This implies that for certain initial conditions the quantum evolution 
is the same for both positive and negative $U$ \cite{41}.

\section{DISCUSSION}

\medskip
We have investigated numerically 2D and 3D models of non-interacting 
fermions and 1D model of two interacting fermions, all cases with
random nearest-neighbour hopping in bipartite systems.
They belong to the chiral orthogonal universality 
class where disorder respects chiral  symmetry, manifesting 
itself in spectral singularities and certain anomalous 
localization behavior at $E=0$. 
It should be pointed out that the effect of chiral symmetry 
in disordered systems was suspected more than two decades ago 
in studies of off-diagonal disorder \cite{8} 
and  about a decade ago in similar studies of BdG
Hamiltonians which describe dirty superconductors \cite{42}.
It was believed that exceptions from the scaling theory
of localization might arise in these systems, 
such as spectral singularities and
a kind of delocalization at the band center in an otherwise localized spectrum. 
Moreover, depending on each author's viewpoint 
the critical behavior of the undoubtedly critical multifractal state
at $E=0$ was named either delocalization \cite{19} or power-law localization 
\cite{13}, although both meant weakening of localization close to $E=0$
where the localization length diverges.
The $E=0$ critical state can be fully described by multifractal
exponents via scaling the amplitude moments and its fractal dimensions,
such as $D_{2}$, vary strongly with the strength of off-diagonal disorder 
ranging from space filling extended states $D_{2}\to d$ for weak disorder
to point-like localized states $D_{2}\to 0$ for strong disorder. 
This transition in the multifractal spectrum 
is known as a ``freezing transition" \cite{20}. 
Below ``freezing" computed mean and typical values agree 
but not above ``freezing" where the means are determined by the tails 
of the corresponding distributions.
In 3D for weak off-diagonal disorder 
where most states with $E\ne 0$ are extended the critical $E=0$ state 
also behaves as an extended state with $D_{2}\approx d$.
This picture also agrees with early transmission studies 
for particle-hole symmetric BdG Hamiltonians which showed 
that quasi-extended states might exist \cite{42} or not {\cite{43}, 
depending on the chosen energy $E$. The present work  summarizes 
all effects of chiral symmetry such as spectral singularities in 2D 
for zero mean off-diagonal disorder which become strong $1/E$ Dyson-type 
singularities for strong off-diagonal disorder, 
while the corresponding  level-statistics 
varies around the critical semi-Poisson 
curve for finite systems of size smaller or equal
than the localization length
($L\lesssim \xi$) suggesting that the $E\ne 0$ 
states from multifractal become fully localized 
for infinite size except the chiral $E=0$ energy state which is
multifractal for any size, disorder and dimension \cite{44,45}.
For  strong off-diagonal disorder we find that the behavior for the rest 
of $E\ne 0$ states is indistinguishable from that of ordinary  
2D disordered systems with the appropriate localization length.

\medskip
We find that the chiral symmetry realized in bipartite 
lattices when the disorder connects one sublattice to the other 
is characterized by enhanced dos and weakening of localization with 
a critical state at $E=0$, having different
spectral and localization properties from the $E\ne 0$ states.
This answers questions (i) and (ii) set in the introduction.
Concerning question (iii) 2D delocalization in this
context seems to require interacting electrons and possibly
broken chiral symmetry \cite{29}.
About question (iv), the effect of chiral disorder on the 
electron-electron interaction for two particles seems to be
no different than that of diagonal disorder. The main message
of section V is that off-diagonal disorder has the same effect
as diagonal disorder for coherent two interacting particle propagation, 
with the presence of ultraslow diffusion. 
Our study also offers answers to questions 
about the eigenstates. Outside the middle of the band ($E\ne 0$)
criticality is expected to occur only for finite systems\cite{36}.
Instead, the $E=0$  state is critical with fractal exponents 
close to space filling for small $W$ (extended states) 
and point-like for large $W$ (localized states) \cite{40}.
This is  similar to what one expects for ordinary Anderson localized states 
with diagonal disorder. 

\medskip
In summary,  we have presented numerical
results for the density of states
with off-diagonal disorder in order to test various theories 
which describe the observed spectral singularities,
from ${\it Dyson}$ to ${\it Gade}$ and ${\it Motrunich}$ 
${\it {etal}}$ forms. 
These results complement many analytical methods 
developed especially for unitary chiral systems.
For ${\it {zero}}$ ${\it mean}$
off-diagonal disorder our main finding is that the form of the 
spectral singularity in 2D is stronger than the log-type 
singularity of the pure system and can be fitted 
either with a weak power-law or by the intermediate 
Gade or Motrunich ${\it {etal}}$ singularities and also 
their formula but with a new exponent $\kappa$. 
In 3D the dos is a little enhanced at $E=0$ but when fitted 
to a power-law the obtained divergence, if any, is very weak. 
The corresponding level-statistics for finite systems in 2D is close 
to the critical semi-Poisson curve and in 3D the Wigner curve,
implying critical and extended states for the majority of states 
in the band for 2D, 3D, respectively. 
For logarithmically ${\it {strong}}$ off-diagonal disorder our results are 
indistinguishable from ordinary strongly disordered systems except 
at the midband critical state $E=0$ where the chiral symmetry 
is responsible for anomalous delocalization.  
Emphasis is given 
in 2D where exceptions from generic localization are more pronounced 
in the form of large localization lengths and multifractality of all 
states for sizes below their localization length. 
However, for large scales a chiral symmetric disordered system behaves like 
an ordinary disorder system with only two visible differences: (i) Spectral 
singularities at the chiral energy $E=0$ which become $1/|E|$ Dyson-like 
for strong disorder or low dimensionality and (ii) the presence of 
multifractality at the critical $E=0$ symmetry point which becomes 
close to extended for zero mean disorder and localized for strong disorder. 
At other energies the chiral disorder has the same effect as 
ordinary disorder.

\medskip
* sevagel@cc.uoi.gr

\medskip
${\bf Acknowledgements:}$ We would like to thank the referees for
useful comments.

\end{document}